\documentclass[fleqn,usenatbib]{mnras}

\usepackage{newtxtext,newtxmath}

\usepackage[T1]{fontenc}
\usepackage{ae,aecompl}

\usepackage{graphicx}	
\usepackage{amsmath}	
\usepackage{amssymb}	

\newcommand{\logg}{\mbox{$\log g$}}

\newcommand{\Liouv}{{\sl Liouville}}

\newcommand{\glh}{{\sl GALAH}}

\newcommand{\ktwohermes}{{\sl K2-HERMES}}

\newcommand{\tesshermes}{{\sl TESS-HERMES}}

\newcommand{\esagaia}{{\sl Gaia}}
\newcommand{\gaia}{{\sl Gaia DR2}}
\newcommand{\tgas}{{\sl Gaia-TGAS}}

\newcommand{\hipparcos}{{\sl Hipparcos}}

\newcommand{\rvs}{{\sl Gaia DR2 RVS}}
\newcommand{\rvsfull}{{\sl Radial Velocity Spectrometer}}

\def\vphi{\ifmmode{\>V_{\mathrm{\phi}}}\else{$V_{\mathrm{\phi}}$}\fi}

\newcommand{\vphivR}{$(V_R,V_\phi)$}
\newcommand{\vphiR}{$(R,V_\phi)$}
\newcommand{\xy}{$(X,Y)$}
\newcommand{\zVz}{$(z,V_{\rm z})$}
\newcommand{\mnvR}{$\langle V_R \rangle$}
\newcommand{\mnvz}{$\langle V_z \rangle$}

\newcommand{\kms}{{\rm km s}$^{-1}$}
\newcommand{\angz}{\rm $L_{\rm Z}$}
\newcommand{\energy}{$(E - E_{\rm circ}(R_{\odot}))/V_{\rm circ}^{2} (R_{\odot})$}

\newcommand{\msun}{\rm $M_{\odot}$}

\title[Linking ridges, arches and vertical waves]{The GALAH survey and \gaia{}: Linking ridges, arches and vertical waves in the kinematics of the Milky Way}

\author[Shourya Khanna et al.]{
Shourya Khanna$^{1,2}$\thanks{E-mail: skha2680@uni.sydney.edu.au},
Sanjib Sharma$^{1,2}$,
Thor Tepper-Garcia$^{1,2}$,
Joss Bland-Hawthorn$^{1,2,3}$,
\newauthor Michael Hayden$^{1,2}$,
Martin Asplund$^{6,2}$, 
Sven Buder$^{7}$,
Boquan Chen$^{1,2}$,
\newauthor Gayandhi M. De Silva$^{5,1}$,
Ken C. Freeman$^{6}$,
Janez Kos$^{4,1}$,
Geraint F. Lewis$^{1}$,
Jane Lin$^{6}$,
\newauthor Sarah L. Martell$^{8,2}$,
Jeffrey D. Simpson$^{8}$, 
Thomas Nordlander$^{6,2}$,
Dennis Stello$^{8}$,
\newauthor Yuan-Sen Ting$^{9,10,11}$,
Daniel B. Zucker$^{5}$,
Toma\v{z} Zwitter$^{4}$
\\
$^{1}$Sydney Institute for Astronomy, School of Physics, A28, The University of Sydney, NSW, 2006, Australia\\
$^{2}$ARC Centre of Excellence for All Sky Astrophysics in Three Dimensions (ASTRO-3D)\\
$^{3}$Miller Professor, Miller Institute, UC Berkeley, Berkeley CA 94720\\
$^{4}$Faculty of mathematics and physics, University of Ljubljana, Jadranska 19, 1000 Ljubljana, Slovenia \\
$^{5}$Department of Physics and Astronomy, Macquarie University, Sydney, NSW 2109, Australia\\
$^{6}$Research School of Astronomy \& Astrophysics, Australian National University, ACT 2611, Australia\\
$^{7}$Max Planck Institute for Astronomy (MPIA), Koenigstuhl 17, D-69117 Heidelberg\\
$^{8}$School of Physics, University of New South Wales, NSW 2052, Australia\\
$^{9}$Institute for Advanced Study, Princeton, NJ 08540, USA\\
$^{10}$Department of Astrophysical Sciences, Princeton University, Princeton, NJ 08544, USA\\
$^{11}$Observatories  of  the  Carnegie  Institution  of  Washington,  813  Santa Barbara Street, Pasadena, CA 91101, US\\
}

\date{Accepted XXX. Received YYY; in original form ZZZ}

\pubyear{2019}
\begin{document}
\label{firstpage}
\pagerange{\pageref{firstpage}--\pageref{lastpage}}
\maketitle

\begin{abstract}
\gaia{} has revealed new small-scale and large-scale patterns in the phase-space distribution of stars in the Milky Way. In cylindrical Galactic coordinates $(R,\phi,z)$, ridge-like structures can be seen in the \vphiR{} plane and asymmetric arch-like structures in the \vphivR{} plane. We show that the ridges are also clearly present when the third dimension of the \vphiR{} plane is represented by $\langle z \rangle$, $\langle V_z \rangle$, $\langle V_R \rangle$, $\langle$[Fe/H]$\rangle$ and $\langle[\alpha/{\rm Fe}]\rangle$. The maps suggest that stars along the ridges lie preferentially close to the Galactic midplane ($|z|<0.2$ kpc), and have metallicity and 
$\alpha$ elemental abundance similar to that of the Sun. We show that phase mixing of disrupting spiral arms can generate both the ridges and the arches. It also generates discrete groupings in orbital energy $-$ the ridges and arches are simply surfaces of constant energy. We identify 8 distinct ridges in the \gaia{} data: six of them  have constant energy while two have constant angular momentum. Given that the signature is strongest for stars close to the plane,
the presence of ridges in $\langle z \rangle$ and $\langle V_z \rangle$ suggests a coupling between planar and vertical directions. We demonstrate, using N-body simulations that such coupling can be generated both in isolated discs and in discs perturbed by an orbiting satellite like the Sagittarius dwarf galaxy.  
\end{abstract}





\section{Introduction}
\label{introduction}

\graphicspath{{figures/}} 
\begin{figure*}
\includegraphics[width=2.0\columnwidth]{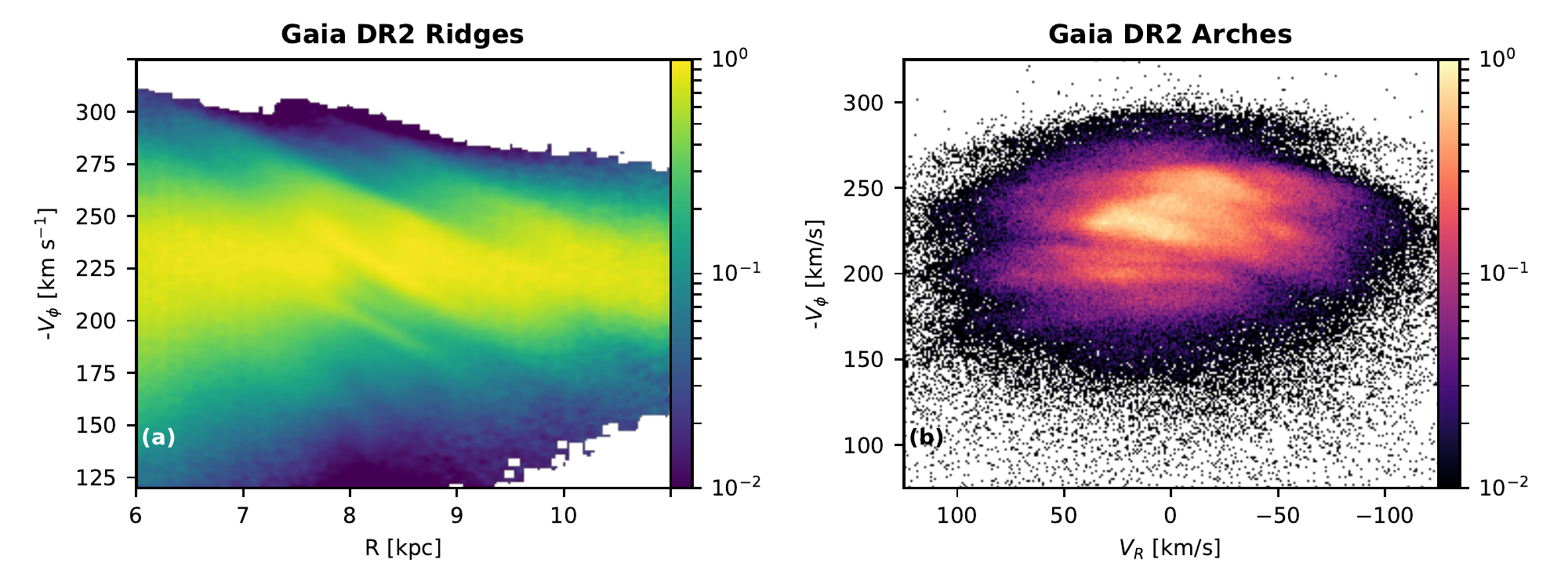}
\caption{New substructure revealed by \citet{2018Natur.561..360A} using \gaia{}. a) The \vphiR{} plane shows diagonal substructures called `Ridges'. Stars are selected to lie in $(|\phi-180.0^{\circ}|<25^{\circ})\&(|R-R_{\odot}|/{\rm kpc}<3.5)$. b) The \vphivR{} plane shows curved substructures called `Arches'. Stars are selected to lie in $(|\phi-180.0^{\circ}|<25^{\circ})\&(|R-R_{\odot}|/{\rm kpc}<0.25)$. The colorbars represent logarithmic density. \label{fig:ridge_arch_gaia}}
\end{figure*}
The second data release of the \esagaia{} astrometric mission \citep[\textit{DR2}: ][]{2018A&A...616A...1G} has heralded a new era in the field of Galactic dynamics. The rich dataset provides position, parallax and proper motion ($\alpha,\delta,\omega,\mu_{\alpha},\mu_{\delta}$) for over a billion stars at unprecedented precision (e.g., $\sigma_{\mu}\approx10$ micro-arcseconds yr$^{-1}$ for $G<14$). In addition, a subset of the full dataset includes line-of-sight velocities from the \rvsfull{} \citep[\rvs{},][]{2018A&A...616A...7S} for about 7 million stars, thus providing full 6D phase-space information for this sample. The sheer number of objects covered by \gaia{}, combined with its high precision, allows us to map the Galaxy's kinematics in a volume more than an order of magnitude larger than that covered by \hipparcos{} \citep{1997A&A...323L..49P}.  

The revelation of multiple new stellar streams \citep{2018MNRAS.481.3442M,2018ApJ...863L..20P}, evidence of non-axisymmetry through substructure in velocity \citep[e.g.,][]{2018A&A...616A..11G,2019MNRAS.484.3291T}, inter alia, have all helped build a good consensus that the Galactic disc is far from being in dynamic equilibrium.
Particularly remarkable was the discovery made by \citet[][A18 hereafter]{2018Natur.561..360A}, which revealed a spiral pattern in the \zVz{} plane density around the Solar neighbourhood, the
so-called phase-spiral\footnote{This feature has been variously referred to as the phase-plane spiral \citep[][]{bin18a} and the phase-space spiral \citep[e.g.][]{2018arXiv181109205K}. Consistent with the traditional use of ``phase mixing'' rather than ``phase-space mixing,'' we adopt the more compact language of phase-spiral \citep[e.g.][]{2019MNRAS.tmp..222B}. The distinctive phase pattern has also been described as the ``snail'' or ``snail shell'' (A18).} (their Fig. 1). The phase-spiral, seen most strongly when color coded by \vphi{}, is thought to be a signature of the Galaxy relaxing from a disturbed state, through phase-mixing. Using toy models, both A18 and \citet{bin18a} suggested that the phase-spiral was evidence of the Galaxy's interaction with the Sagittarius dwarf galaxy (Sgr), and further constrain the last impact to about 0.5 Gyr ago. Recent N-body simulations \citep[e.g., by][]{2019MNRAS.tmp..222B,2019MNRAS.485.3134L}  have shown that tidal interaction with Sgr can indeed reproduce the phase-spiral seen in \gaia{}, and suggest a similar or younger timescale for the interaction. However, \cite{2018arXiv181109205K} have shown that the phase-spirals can also be generated through an entirely internal mechanism. In their simulations, they show that the buckling of the Galactic bar can generate bending waves in the disc. This is able to create the phase-spiral, and the wave takes about 0.5 Gyr to travel to the outer disc ($\sim$10 kpc). The spirals survive well after the end of the buckling phase, where these bending waves are supported by the disc self-gravity. These results show that it is non-trivial to distinguish between an internal or an external perturbation.

A18 also revealed that the \vphiR{} space has substructure in the form of diagonal ridges (\autoref{fig:ridge_arch_gaia}a) and that the \vphivR{} space has arches (\autoref{fig:ridge_arch_gaia}b), some of which are asymmetric about the $V_R=0$ line. They suggest that arches are just the projection of ridges in the \vphivR{} velocity space; however, ridges can also be present without any arches. \cite{2018A&A...619A..72R} identified some of the \vphivR{} arches and traced their median $V_{\phi}$ at different Galactocentric radii $R$, suggesting that the arches and ridges are linked. 

We show that physical understanding of the connection between the ridges and arches is still missing. Moreover, any connection between these dynamical excitations and the phase-spiral has yet to be clearly established. All of these phenomena have distinct properties (amplitudes, wavelengths, etc.) and their unification is the topic of a later paper.

\begin{table}
\centering
\caption{Data quality cuts on \glh{} DR2.\label{tab:galah_flags}}
\begin{tabular}{ll}
\hline
\hline
Selection & Comments  \\
\hline
$9 < V_{JK} < 14 $ & - \\
0$\leq$ Field id $<7339$ & Excludes data without proper selection\\
& function \\
\hline
\end{tabular}
\end{table}

Several models using various simulation techniques have been proposed to explain the ridges and arches. Most models explain either the ridges or the arches, but not necessarily both. Resonant scattering by non-axisymmetric features rotating with a fixed pattern speed, e.g., the bar or spiral arms, has been shown to generate arches.   
\cite{2000AJ....119..800D} showed that one prominent asymmetric arch and two other weak arches can be developed by a bar, which has since been demonstrated by several other simulations  
\citep[e.g., ][]{2014A&A...563A..60A,2017MNRAS.471.4314M,2018MNRAS.477.3945H,2019MNRAS.484.4540H,2017ApJ...840L...2P}. 
A18 further showed that resonance with a bar can also generate ridges, but only one or possibly two ridges can be seen in the solar neighborhood as compared to the many seen in \gaia{}. More recently, simulations by \cite{2019arXiv190107568F} also showed that the outer Lindblad resonance of the Galactic bar could give rise to one of the prominent ridges in the \vphiR{} plane and a Hercules-like feature in the \vphivR{} plane. 

Phase-mixing models have also been used to explain these kinematic features. In such models, test particle simulations are employed. Test particles are set up to mimic a perturbation and are then evolved in a Milky Way-like potential.
A18 showed that ridges can be generated using a horizontal phase-mixing model, but did not show if they lead to arches. Moreover, the physical motivation for the model was also not made clear. 
\cite{2009MNRAS.396L..56M} showed that phase wrapping after interaction with a dwarf galaxy can produce multiple arches in the \vphivR{} plane, similar to those seen in the solar neighbourhood \citep[see also][]{2012MNRAS.423.3727G}, but they do not explore the occurrence of ridges. 

This raises an interesting question: {\it is the impact with a dwarf galaxy necessary to see multiple ridges?} \cite{2018MNRAS.480.3132Q} point out that the arches seen in \gaia{} are tilted about the $V_{R}=0$ line, but those generated  by the phase-wrapping model of \cite{2009MNRAS.396L..56M} are symmetrical. They propose that a model in which the stars that have recently crossed spiral arms at their apocenter or pericenter can explain the asymmetric arches; however, they do not study the ridges. 

\citet{2018MNRAS.481.3794H} consider a potential with 2D transient spiral arms that wind up over time, and using the backward integration technique of \citet{2000AJ....119..800D}, show that this perturbation can give rise to features such as the Hercules stream in the \vphivR{} plane, as well as multiple ridges in the \vphiR{} plane, and multiple asymmetric arches. Transient spiral arms have been shown to develop in self-gravitating disc simulations \citep{2011MNRAS.410.1637S}. This sets up the motivation to look for ridges and arches in simulations of this kind. 
\citet{2011MNRAS.417..762Q} showed that asymmetric arches can be generated in self-gravitating N-body simulations, but did not study the ridges in \vphiR{}. 
\cite{2019MNRAS.485.3134L} studied N-body simulations involving interaction with a dwarf galaxy, and were able to generate ridges in \vphiR{}, but only one arch or moving group could be seen in the \vphivR{} plane. An interesting question to ask is whether the source of the ridges and arches is internal or external, and how we could distinguish between the two. Can a phase-mixing model motivated by transient spiral arms explain multiple ridges and multiple asymmetrical arches?

Vertical waves have also been reported in the \gaia{} data \citep[e.g., ][]{2018A&A...616A..11G,2019MNRAS.482.1417B}. Already with the limited coverage of \tgas{}, \citet{2018MNRAS.478.3809S} and \cite{2018ApJ...864..129H} found that the vertical velocity in the solar neighborhood varies with angular momentum $J_{\phi}$. They found a large-scale trend of $V_z$ increasing monotonically with $J_{\phi}$, which is a signature of the Galactic warp. Superimposed on
this large-scale trend, they also found undulations (or corrugations) indicative of a wave-like pattern. Undulations in the profile of $V_z$ as a function of Galactocentric radius $R$ were also reported by \cite{2018MNRAS.479L.108K}. \cite{2013MNRAS.429..159G} and \citet{2016ApJ...823....4D} both show that undulations in the $V_z(R)$ profile can be seen in N-body simulations involving interaction with Sgr. However, the variation of $V_z$ as a function of angular momentum was not studied. Are these vertical waves linked to ridges and arches? Can these vertical waves be seen in simulations with or without the interaction of Sgr? This is a question we attempt to address.

In this paper, we revisit the \vphiR{} ridges seen in \gaia{}. First, we dissect and characterize the ridges using radial velocity, vertical height, and vertical velocity.  
Furthermore, we explore the nature of ridge stars by considering elemental abundances from \glh{} and relate this to the nature of the perturbation itself. Next, we simulate phase mixing of spiral arms and show how this model can be used to understand the connection between the \vphiR{} ridges and the \vphivR{} arches. Finally, we carry out N-body simulations of the Galactic disc, both with a Sgr-like perturber and without any perturber and study the phase-space features in these simulations and compare them with those seen in \gaia{}.

\section{Data set and methods}
\label{sec:dataset}

Throughout the paper, we adopt a right-handed coordinate frame in which the Sun is at a distance of $R_{\odot}=8.2$ kpc from the Galactic center \citep{2016ARA&A..54..529B}, consistent with the new ESO Gravity measurement \citep{2018A&A...615L..15G}, and has Galactocentric coordinates $(X,Y,Z) = (-8.2,0,0.25)$ kpc. 
The cylindrical coordinate angle $\phi={\rm tan}^{-1}(Y/X)$ increases in the anti-clockwise direction, while the rotation of the Galaxy is clockwise. The heliocentric Cartesian frame is related to Galactocentric by $X_{\rm hc}=X+R_{\odot}$, $Y_{\rm hc}=Y$ and $Z_{\rm hc}=Z$. $X_{\rm hc}$ is negative toward $\ell=180^\circ$ and $Y_{\rm hc}$ is positive towards Galactic rotation. For transforming velocities between heliocentric and Galactocentric frames we use $(\dot{X}_{\odot},\dot{Y}_{\odot},\dot{Z}_{\odot})=(U_{\odot},\Omega_{\odot}R_{\odot},W_{\odot})$.
Following \cite{2010MNRAS.403.1829S}, we adopt $(U,V,W)_{\odot}=(11.1,12.24,7.25)$ \kms{}, while for the azimuthal component we use the constraint of  $\Omega_{\odot}=30.24$ \kms{}kpc$^{-1}$ which is set by the proper motion of Sgr A*, i.e., the Sun's angular velocity around the Galactic center \citep{2004ApJ...616..872R}. This sets the rotation velocity at the Sun to $V_{\phi,\odot}=-248$ \kms{}, and thus the circular velocity at the Sun to $V_{c, \odot}=-236$ \kms{}. 
We now describe the astrometric and spectroscopic data that we use in this work and the quality cuts that we apply on them.

\subsection{\textit{Gaia DR2} RVS sample}
\label{sec:rvs_sample}
In this paper we make use of the \gaia{} 
radial velocity sample (\rvs) which provides full 6D phase space information ($\alpha, \delta, \omega, \mu_{\alpha}, \mu_{\delta},V_{\rm los}$). We selected stars with positive parallax and with parallax precision $\sigma_{\omega}/\omega <$ 0.2, which gave a sample of $6376803$ stars. The SQL query used to generate the sample is given in \autoref{app:gaia_sql}. 
We estimated distance as $1/\omega$, which is reasonably accurate for our selected stars and for the purpose of this paper \citep{2018A&A...616A...9L}.

\subsection{GALAH DR2 sample with elemental abundances}
\label{sec:galah_sample}
The spectroscopic data used here is taken from an internal release of \glh{} \textit{DR2}, which includes, public data \citep[][\textit{DR2}]{2018MNRAS.478.4513B}, and fields observed as part of the \ktwohermes{} \citep{2018AJ....155...84W} and \tesshermes{} \citep{2018MNRAS.473.2004S} programs. To maintain the survey selection function, we have applied the quality cuts summarised in \autoref{tab:galah_flags}, which gives a total of 465870 stars cross-matched with \gaia{}. This internal release includes non-LTE corrections on [Fe/H] but not on [$\alpha$/Fe]. For the kinematics of this dataset, we make use of the parallax and proper motion ($\omega,\mu_{\alpha}, \mu_{\delta}$) from \gaia{}, but use the highly precise radial velocities from \glh{}, which have typical error of 0.1 \kms{} \citep{2018MNRAS.481..645Z}. Since we are mainly interested in nearby stars, we restrict our \glh{} sample only to dwarfs, by applying a surface gravity cut of (\logg $> 3$), which results in a final sample of 258289 stars. This avoids any issues related to systematic errors in stellar parameters between dwarfs and giants.

\begin{table}
\begin{center}
\caption{Parameters for the isolated Galaxy (Model P). Column headers are as follows: $M_{\rm t}$ := total mass ($10^{9}$ \msun); $r_{\rm s}$ := scalelength (kpc); $r_{\rm tr}$ :=  truncation radius (kpc); $N_{\rm p}$ := number of particles ($10^{6}$). }
\label{tab:Galaxy}
\begin{tabular}{lccccc}
\hline
\hline
 				& Profile	& $M_{\rm t}$ 	&  $r_{\rm s}$ 		& $r_{\rm tr}$ 	& $N_{\rm p}$  ~\\
\hline
Galaxy\\
\hline
DM halo			& H			& $10^3$		& 38.4		& 250	& 	10	~\\
Bulge			& H			& 9				& 0.7		& 4		&	1	~\\
Thick disc		& MN		& 20			& 5.0$^{a}$	& 20	&	2	~\\
Thin disc		& Exp/Sech	& 28			& 3.0$^{b}$	& 20	&	3 ~\\
\end{tabular}
\end{center}
\begin{list}{}{}
\item {\em Notes}. H := \citet{her90a} profile; MN := \citet{miy75a} profile; Exp := radial exponential profile.; Sech := vertical ${\rm sech^2} z$ profile.\\
$^{a}$scaleheight set to 0.5 kpc.\\
$^{b}$scaleheight set to 0.3 kpc.
 \end{list}
\end{table}

\begin{table}
\begin{center}
\caption{N-body models and  properties of the perturber \label{tab:perturber}. Column headers are as follows-- $M_{\rm
tot}$: total mass ($10^{9}$ \msun); $M_{\rm tid}$ : tidal mass ($10^{9}$
\msun); $r_{\rm tr}$ : truncation radius of dark matter in kpc; $N_{\rm p}$ : number of
particles ($10^{5}$ and $v_0$: approximate initial orbital speed in \kms. See the notes below the table for more information.}
\begin{tabular}{lccccc}
\hline
\hline
Model						& $M_{\rm tot}$ 	& $M_{\rm tid}$ 	& $r_{\rm tr}$ 	& $N_{\rm p}$  & $v_0$~\\
\hline
P (unperturbed/isolated galaxy) &  0				& 				& 			& 				& 	\\
S (intermed. mass, one transit) &  50				& 30				& 19			& 1				& 360	\\
R (high mass, one transit)      & 100				& 60				& 24			& 2				& 372	\\
\end{tabular}
\end{center}
\begin{list}{}{}
\item {\em Notes}. Both the DM halo and the stellar component are initially modelled as Hernquist spheres with $r_{\rm s} = 9.8$ kpc and $r_{\rm s} = 0.85$ kpc, respectively. The mass of the stellar component is $4 \times 10^8$ \msun\ in either case, split among $4\times10^4$ particles.
\end{list}
\end{table}

\subsection{Phase mixing simulations}
\label{sec:toy_model}
To understand the origin of the phase-space substructures like ridges and arches, we perform simulations in which 
spiral arms phase mix and disrupt over time. The simulations are motivated by the desire to mimic the effect of transient spiral arms.
For this we consider an initial distribution of particles confined to four thin spiral arms. The $i^{th}$ arm is setup as an \textit{Archimedean} spiral, with azimuth:
\begin{equation}
\phi = \frac{1}{b}(r - a) + i\frac{\pi}{2},
\end{equation} 
where, $0 < a < 2\pi$ controls the orientation of the spiral, $b=\pi/10$ controls the tightness of the winding, and $r = R/8$ kpc. The radial distribution was assumed to be skew normal with skewness of 10, location parameter of 4 and scale parameter of 6. This is to ensure that there are enough particles in the Solar neighbourhood-like volume. The radial velocity was sampled from $\mathcal{N}(0,20)$ and the azimuthal velocity from $\mathcal{N}(\Theta(R),20)$, where $\Theta(R)$ denotes the circular velocity. For simplicity, the particles were set up in the midplane with zero vertical velocity. A total of 640000 particles were evolved for 650 Myr with {\tt galpy} \citep{2015ApJS..216...29B} using the {\tt MWPotential2014} potential, consisting of an axisymmetric disc, a spherical bulge and a spherical halo. The set up is similar to \cite{2018Natur.561..360A}, but they start with stars arranged in a single line as compared to four spiral arms used by us. A set of movies showing the evolution of the system is available as Supporting Information in the online version of this paper.

\subsection{N-body simulations}
\label{sec:nbody_model}

Another approach that we adopt, in order to gain insight into the origin of phase-space substructures,  is the use of N-body simulations of a
multi-component Galaxy. Ideally, we would like to carry on a detailed modelling of every component of the Galaxy, both collisionless (e.g. dark matter and stars) and gas. This is, however, not feasible as it is both computationally expensive and non-trivial to do. We adopt a common, simplifying assumption: we assume that a pure N-body (rather than a full N-body, hydrodynamical) model is sufficient for our purposes. We caution that neglecting the gas components in these type of experiments may not always be appropriate  \citep[see e.g.][]{tep18b}.

We consider here the following two scenarios as the plausible origin of the ridges: i) instabilities internal to the Galaxy; and ii) tidal (external) interactions. In consequence, we focus our attention on one representative model for each of these. On the one hand, we simulate the evolution of an isolated Galaxy starting from some prescribed initial conditions (see below). On the other hand, we simulate how the stars behave in a Galaxy that has been tidally perturbed by the interaction with a smaller system. It has been suggested that Sgr may lie behind many of the kinematic features revealed by \gaia\ \citep[e.g.][]{2018Natur.561..360A,bin18a,2019MNRAS.485.3134L,2019MNRAS.tmp..222B}. It therefore seems natural in our case to simulate the interaction of the Galaxy with a Sgr-like perturber.

\graphicspath{{figures/}} 
\begin{figure}
\includegraphics[width=1.0\columnwidth]{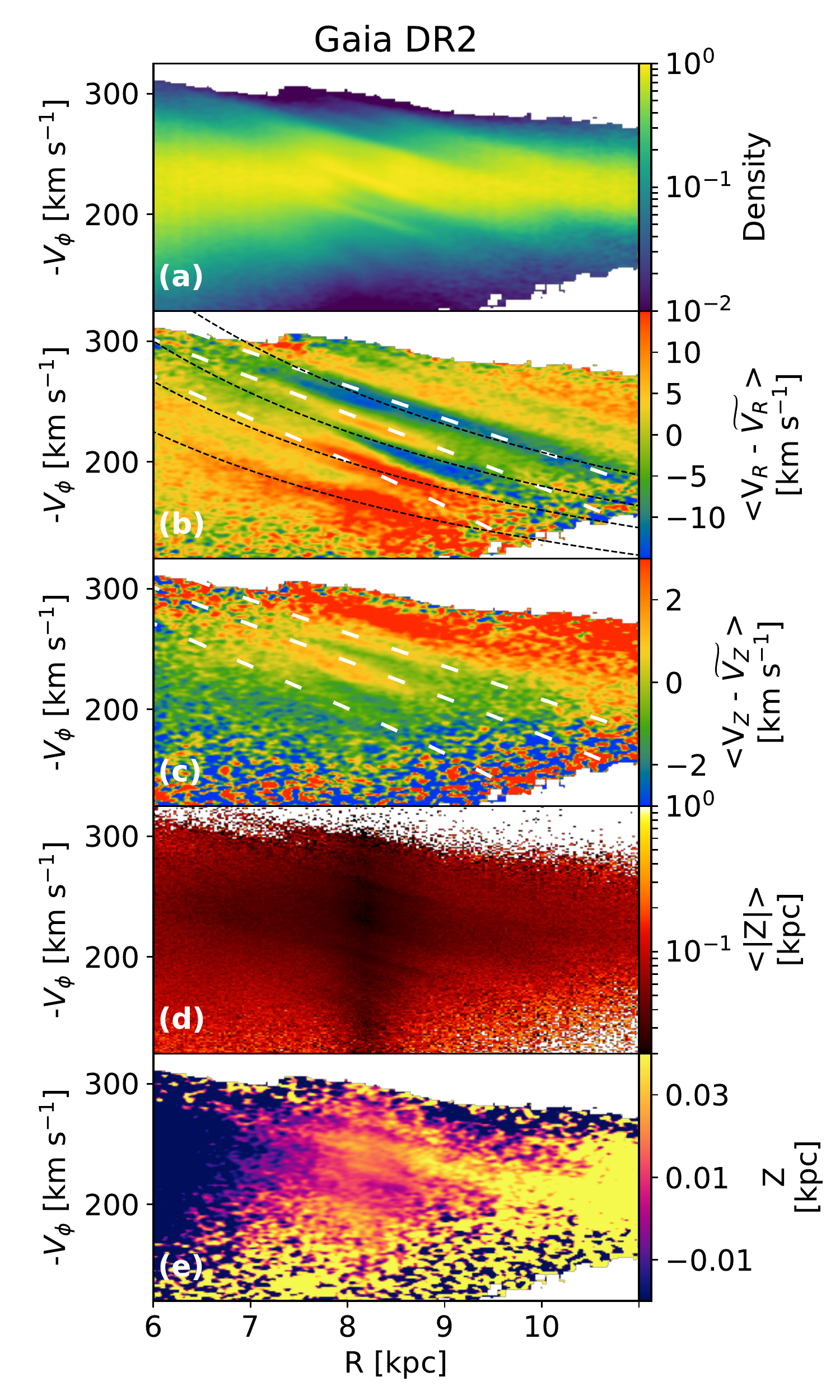} 
\caption{Study of the \vphiR{} plane with \gaia{} RVS. We select stars in the region ($|\phi-180^\circ| < 25 ^\circ$) \& ($|R-R_{\odot}|/{\rm kpc} <3.5$). Heat maps of various quantities are shown; a) probability density of $V_{\phi}$ conditional on $R$ ($p(V_{\phi}|R$)) b)  mean radial motion \mnvR c) mean vertical motion \mnvz{} d) mean absolute distance from the plane $|z|$ and e) distance from the plane z. The white dotted curves represent constant energy for values of -0.112, -0.021, \& 0.097 for \energy{}. Black curves represent constant angular momentum, $L_{\rm z}= (1350,1600,1800,2080)$ kpc \kms{}. \label{fig:vphiR_gaia}}
\end{figure}
\graphicspath{{figures/}} 
\begin{figure}  
\includegraphics[width=\columnwidth]{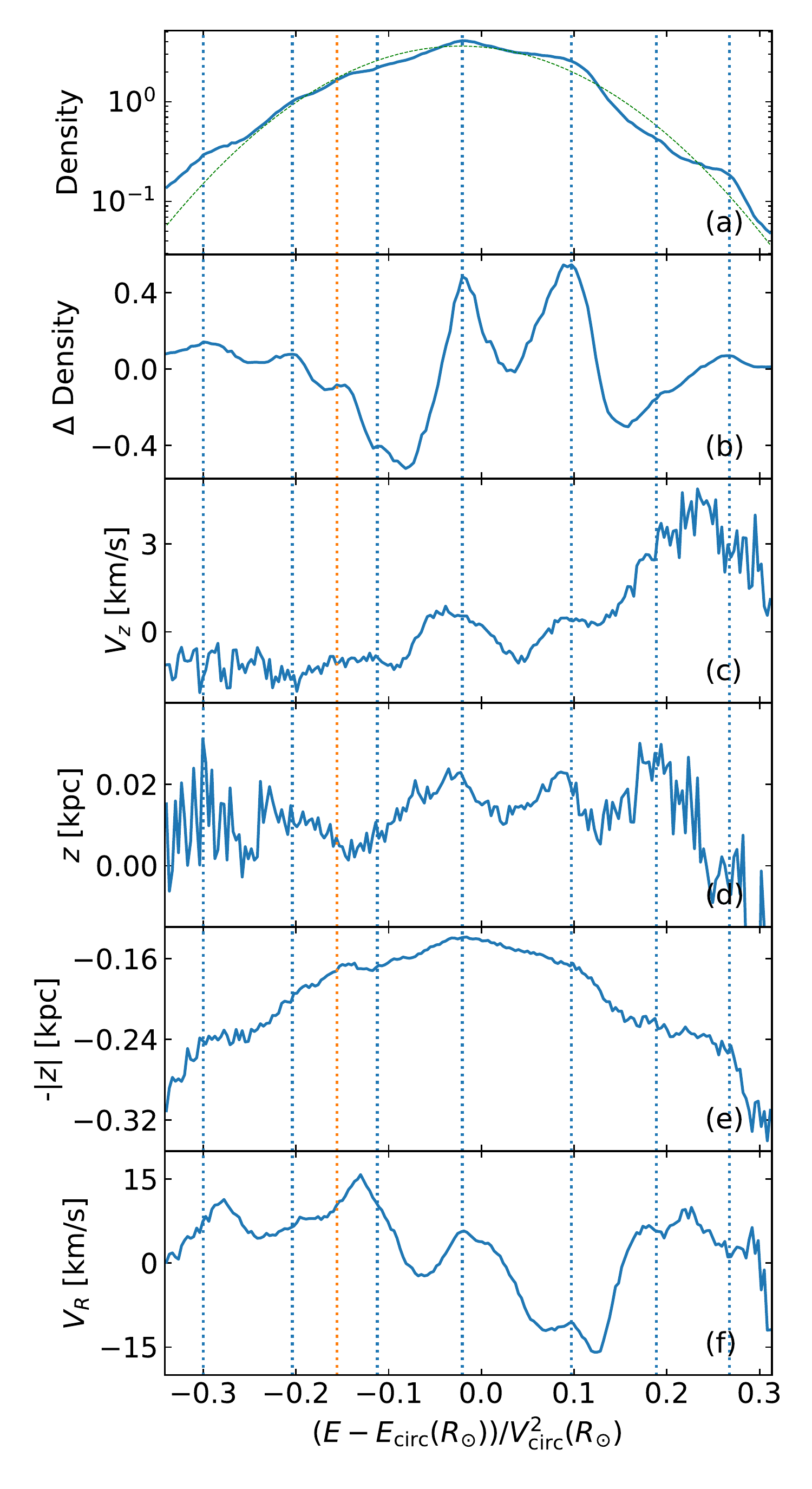}
\caption{Profiles of different quantities as a function of energy for stars in $(|R-R_{\odot}|<1.0)\&(|\phi-\phi_{\odot}|<25)$. a) Density distribution along with best fit skewed normal distribution (green line). b) Residual of density after subtracting a skewed normal distribution. c) Median vertical velocity d) Median vertical distance above the plane. e) Median absolute value of distance from the plane. f) Radial velocity. The vertical dotted lines mark the location of peaks at $[-0.300, -0.204, -0.156, -0.112, -0.021, 0.097, 0.189, 0.267]$. The peaks are approximately regularly spaced with mean separation of 0.095,  except for  the peak at -0.156 corresponding to the Hercules stream. 
\label{fig:gaia_quant_eorb}}
\end{figure}
\graphicspath{{figures/}} 
\begin{figure}
\includegraphics[width=1.0\columnwidth]{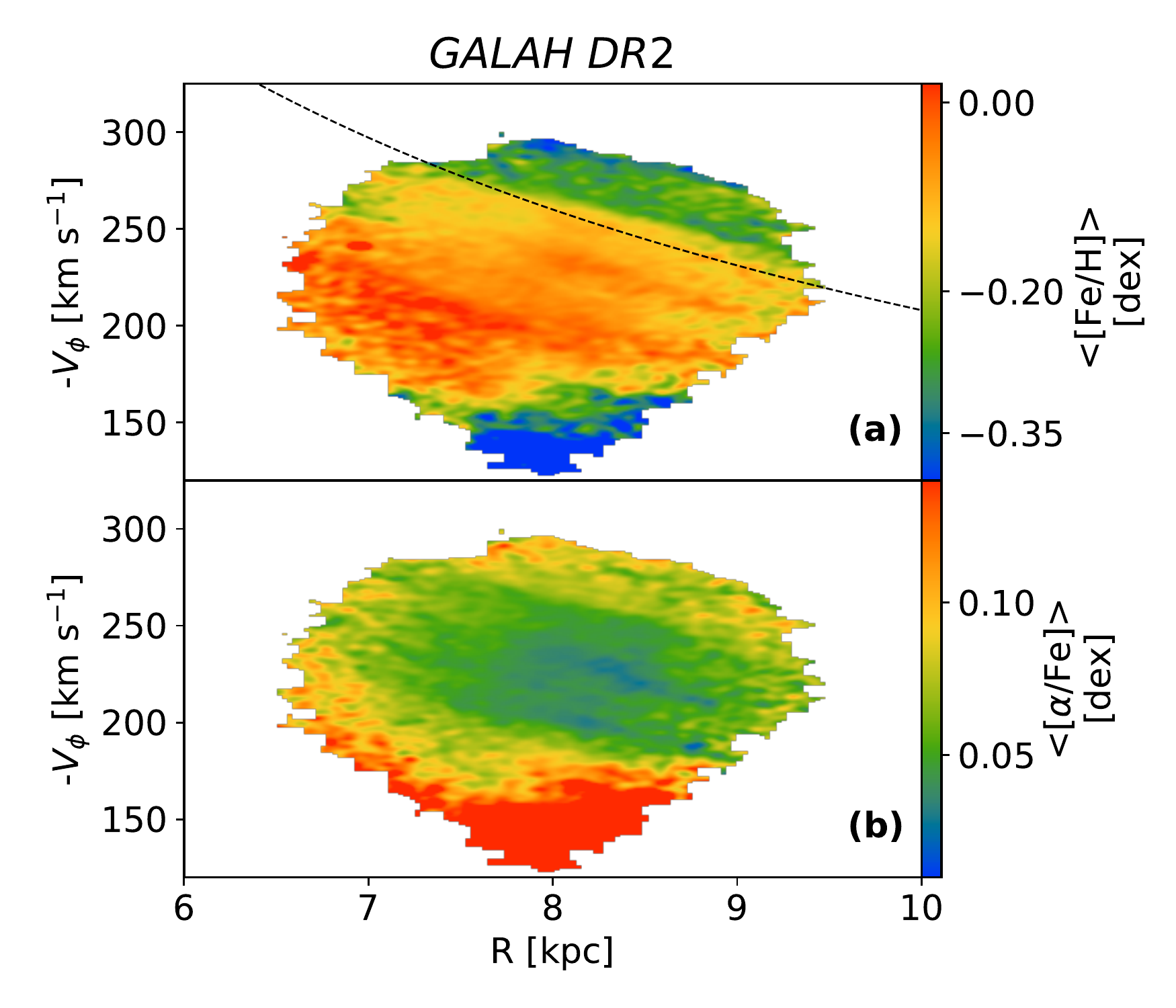}
\caption{Study of the \vphiR{} plane with elemental abundances from \glh{} DR2 for stars in $(|R-R_{\odot}|{\rm /kpc} <3.5)\&(|\phi-\phi_{\odot}|<25)$. (a) [Fe/H] map, with black dotted line at \angz{}$= 2080$ kpc \kms{}, (b) [$\alpha$/Fe] map. Ridges are comprised of metal rich and alpha-poor stars. \label{fig:vphiR_galah}}
\end{figure}

Our isolated model Galaxy consists of four collisionless components: a host DM halo; a stellar bulge; a thick stellar disc; and a thin stellar disc.  We refer to this model as the `isolated' model (Model P); see Table~\ref{tab:Galaxy} for details of this model Galaxy. Note that the values for structural parameters (scalelength, scaleheight, etc.) are only consistent with the range of values inferred from observations \citep[][]{2016ARA&A..54..529B}.   
We assume the distribution of stars in the thick disc is well approximated by a \cite{1975PASJ...27..533M} (MN) profile. This choice, and its corresponding parameter values, are founded on the work of \citet{2014ApJ...794...59K}, who use  precise stellar kinematic information to infer the mass distribution of the Milky Way, assuming that total stellar disc component is well described by a single MN component. They obtain a scalelength and scaleheight of 5 kpc and 0.5 kpc, respectively. Note that \cite{2016ARA&A..54..529B} provide instead mean value of 2 kpc
and 0.9 kpc, respectively, which is however predicated by the assumption that the mass distribution in the (thick) stellar disc be well described by an exponential profile. Our adopted values are consistent with the fact that the scalelength of a MN may differ by a up to factor of $\approx$2 compared to the scalelength of an equivalent exponential disc \citep{1996MNRAS.281.1027F} and that the exponential scaleheight can differ, potentially by the same factor of 2, from the MN scaleheight and that the exponential scaleheight can differ, potentially by the same factor of 2,from the MN scaleheight   \citep{2015MNRAS.448.2934S}. For the thin disc we adopt a scalelength and a scaleheight of 3 kpc and 0.3 kpc, respectively. These values agree well with the mean values quoted by \cite{2016ARA&A..54..529B}. Our adopted mass for the thin disc is well within the range inferred from observations. The adopted mass for the thick disc is higher than quoted by \cite{2016ARA&A..54..529B}. However, it should be noted that our adopted value corresponds to the mass integrated out to 20 kpc. A lower mass is obtained if truncating the disc at a smaller radius. Nevertheless, the total stellar disc mass is consistent with other estimates \citep[e.g.,][]{2013ApJ...779..115B}.  Overall, our choices of component profiles and the values of their corresponding parameters define a valid model for the Galaxy, comparable to the model successfully adopted by others in numerical studies  \citep[e.g.,][]{2018MNRAS.480.4244C}.

Our interaction models consist essentially of a two-component (DM, stellar spheroid) system orbiting an initially isolated Galaxy along an (unrealistic) hyperbolic orbit. The reason for choosing such an orbit rather than a more realistic orbit for the perturber is that, as we have shown previously \citep[][]{2019MNRAS.tmp..222B}, each passage of Sgr across the Galactic plane washes out the kinematic signatures of its previous crossing, thus limiting the time span available between crossings. In contrast, by adopting a hyperbolic orbit we ensure that Sgr transits the Galactic plane (disc) once only, thus facilitating the analysis of its effect on the Galactic stars.

We consider perturbers with total masses of 5 or $10\times10^{10}$ \msun, spanning the mid-to-high range of plausible Sgr masses at infall \citep[e.g][]{nie10a}. Both the stellar system and the dark halo are modelled as truncated \citet[][]{her90a} spheres. Their scale radii are initially set at $0.85$ and 10 kpc, respectively. The stellar system is initially truncated at 2.5 kpc while the truncation radius of the dark halo is listed in Table~\ref{tab:perturber}. A simulation with each of these masses was started with the perturber at $(x,y,z) = 20.8, 0., 45.5)$ kpc on an orbit of eccentricity $e=1.3$ (hyperbolic) and pericentric distance 10 kpc.\footnote{The exact initial initial velocities for model R and model S was $(v_x, v_y, v_z) = (-267, 0, -260)$ \kms\ and $(v_x, v_y, v_z) = (-258, 0, -251)$ \kms, respectively. }

Two key requirements on these type of simulations, imposed by the exquisite detail on the kinematics of stars revealed by the data, are the mass resolution (or particle number) and the limiting spatial resolution. The latter has to be low enough to allow for a correct simulation of the evolution of the dynamically coldest stellar component within $|z| < 0.2$ kpc. The former needs to be high enough to allow for a dense enough sampling of the \vphiR{}.

We choose values for the particle number and spatial resolution such that we fulfill these requirements while keeping the computational cost of the simulations reasonably low. More specifically, we set the limiting spatial resolution at 30 pc, to be compared with 300 pc, the (initial) scaleheight of the cold stellar disc, which is the smallest length scale in our simulation. The adopted particle number varies from component to component, depending on the total mass of the component and the corresponding particle mass; as per our above discussion, the thick and thin stellar discs have been assigned the absolute highest particle number (see Table \ref{tab:Galaxy}).

The simulations' axisymmetric initial conditions, i.e. the particles' positions and velocities for each component were assigned by the technique of \citet[][]{spr05c} as implemented in the {\sc dice} code \citep[][]{per14c}. In doing so, all the components are intended to be in dynamical equilibrium with the total potential of the compound system. However, in reality they will in general be slightly out of equilibrium \citep[e.g.][]{kaz04a}, and even an isolated Galaxy disc will develop some small-scale structure, such as rings and transient spirals. While usually an unwanted numerical artifact in experiments like ours, here we actually need these instabilities in order to investigate their effect on the stellar phase-space kinematics. In real galaxies, such transient features will develop as well, but for reasons yet to be fully understood.

The evolution of the system\footnote{The {\sc dice} and {\sc Ramses} configuration files used to create our initial conditions and to setup our simulations, respectively, are freely available upon request.} in each case is calculated with the adaptive mesh refinement (AMR) gravito-hydrodynamics code {\sc Ramses} \citep[version 3.0 of the code described by][]{tey02a}. Simulation data are stored at approximately $\Delta \tau=$10 Myr intervals. A set of movies showing the evolution of the system in each model are provided at \href{http://www.physics.usyd.edu.au/~tepper/proj_galah_paper.html}{Gaia-GALAH-phase-spiral}.


\section{Results}

We begin by studying the \vphiR{} plane using the observed data. Next we compare the observed results with predictions from two type of simulations, phase mixing simulation of disrupting spirals and disc N-body simulations. 
This is followed by a study of the $(V_{R},V_{\phi})$ plane and subsequently the $(R,V_{R})$ plane. In both cases we compare the observed results with the predictions from the simulations.

\subsection{Analysis of the \vphiR{} plane using the observed data}
\label{sec:ridge_kin}
\subsubsection{Dissection in kinematics and vertical height with Gaia}
\cite{2018Natur.561..360A} revealed the diagonal ridge-like structures in the \vphiR{} density distribution. In this section, we further explore this plane using kinematics and vertical height. 
In \autoref{fig:vphiR_gaia}, we show results using the \rvs{}  sample. We select stars with ($|\phi-180^\circ| < 25 ^\circ$) \& ($|R-R_{\odot}|/{\rm kpc} <3.5$). 


\graphicspath{{figures/}} 
\begin{figure*}
\includegraphics[width=2.0\columnwidth]{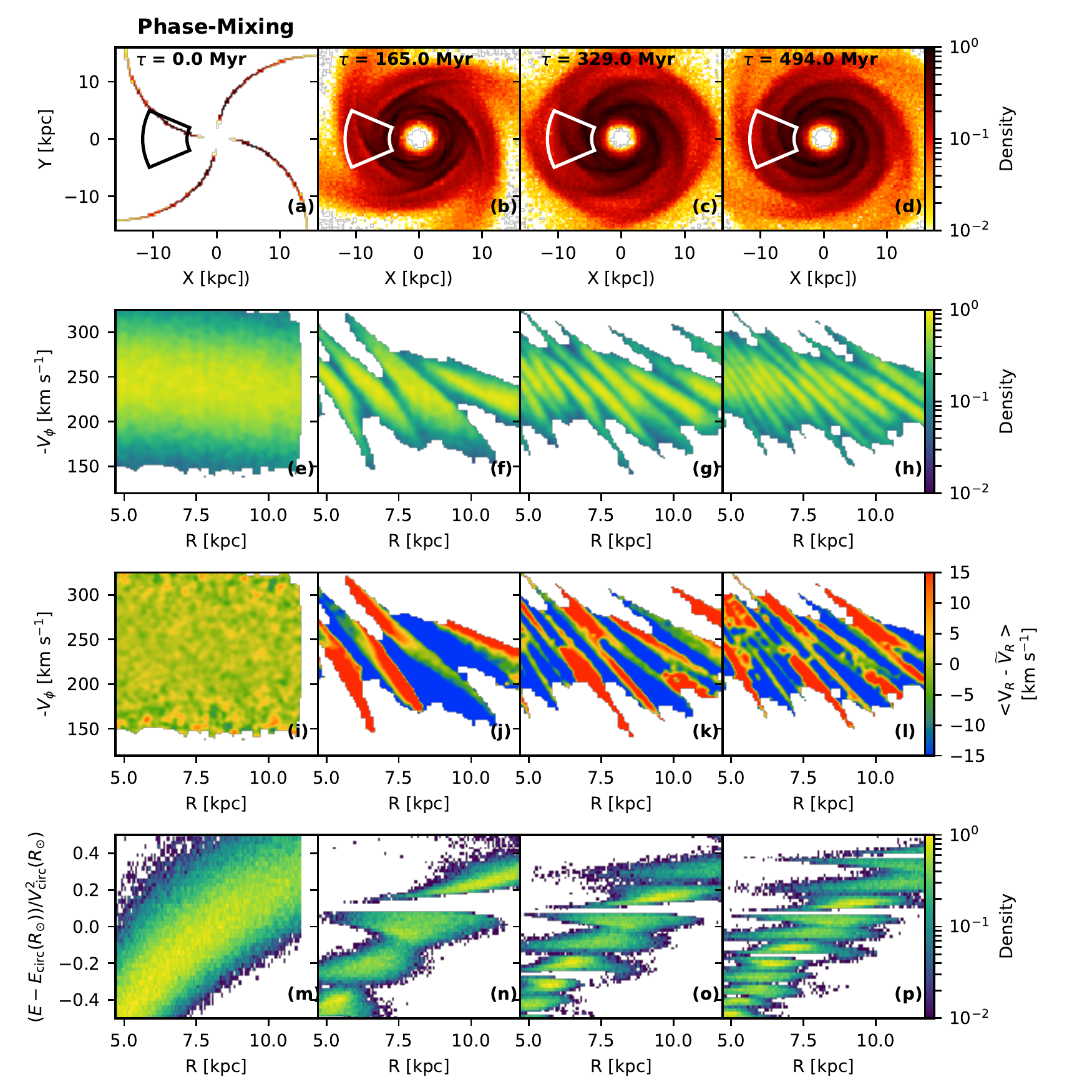}
\caption{Evolution of stars 
in the phase mixing simulation 
(\autoref{sec:toy_model}). Panels (a,b,c,d) show the 
density in $xy$ plane for five different time snapshots. Panels(e-h) show $p(V_{\phi}|R$), the 
probability density of $V_{\phi}$ conditional on $R$. Panels (i,j,k,l) show the \vphiR{} plane mapped by \mnvR{}. Panels (m,n,o,p) show the total energy against $R$. For panels(e to p), stars are selected in $(|R-R_{\odot}|{\rm /kpc}<5.0)\&(|\phi-\phi_{\odot}|<25 ^\circ)$, indicated by the locus in Panels (a-d). \label{fig:phasemix_ridges}}
\end{figure*}


\autoref{fig:vphiR_gaia}a shows the  density distribution \mnvR{} in the \vphiR{} plane. Multiple diagonal ridges are clearly visible, extending between $6<R/{\rm kpc}<12$. Ridges are prominent at $R=8.2$ kpc, but seem to fade away as we move away from the solar radius, this is because, as we move away from the Sun there is a fall in number density of stars and an increase in uncertainty in $R$ and $V_{\phi}$.  
\autoref{fig:vphiR_gaia}b shows a map of \mnvR{} in the \vphiR{} plane. The ridges are more prominent 
and are visible even at large distances from the solar neighborhood. They appear at a similar location to that in the density map in \autoref{fig:vphiR_gaia}a). Stars along the ridges are moving either radially outward or inward, with respect to the background distribution, with \mnvR{}$\approx10$ \kms.

Next we explore the properties of the ridges in the vertical direction.
\autoref{fig:vphiR_gaia}c shows a map of \mnvz{} in the \vphiR{} plane. The ridge structure can again be seen but it is weaker as compared to the \mnvR{} map. Three ridges are clearly visible and the \mnvz{} associated with the structures is about 2 \kms{}.
\autoref{fig:vphiR_gaia}d shows \vphiR{} mapped by $\langle|z|\rangle$, i.e., the mean of the absolute distance from the mid-plane of the disc. The ridges are primarily composed of stars that lie close to the Galactic plane ($|z| \lesssim 0.2$ kpc), as indicated by the distinctive dark color. It is important to note that if stars at all heights above the plane participated in the ridges, the $|z|$ map in \autoref{fig:vphiR_gaia}d would be completely featureless. This preferential distribution must thus be linked to the nature of the perturber responsible for the ridges. Three ridges are also visible in the map of $\langle z \rangle$, i.e. the mean distance from the plane (\autoref{fig:vphiR_gaia}e).

We have overplotted  curves of constant angular momentum (black dotted lines) at $L_{\rm z} = [1350,1600,1800,2080]$ kpc \kms{} and curves of constant orbital energy $E$ (white dashed lines) at \energy{}$=[-0.112, -0.021, 0.097]$. The energy was evaluated using the {\tt MWPotential2014} potential in {\tt galpy} \citep{2015ApJS..216...29B}.
Both curves decrease with $R$ and resemble ridges, and hence either of these physical quantities can be used to label the ridges. 
The difference between the two is that the constant energy curves are straight lines but the angular momenta ones are not. From these plots, it is difficult 
to say if the ridges are constant energy or constant angular momentum, we revisit this issue later in \autoref{sec:toy_model_results}. 

We have shown that the ridges are present in maps of density, kinematics and vertical height. We now investigate if the ridges in the maps of different quantities are correlated with each other. For this we select stars in a narrow range in $R$ and then study the one dimensional profiles of various quantities as a function of orbital energy $E$ (\autoref{fig:gaia_quant_eorb}). We choose $E$ instead of $V_{\phi}$ as ridges are well approximated by curves of constant energy. We use the {\tt MWPotential2014} potential in {\tt galpy} to compute the energy \citep{2015ApJS..216...29B}. Instead of directly using $E$, we use the dimensionless form given by  $E'=$\energy{}, where 
$V_{\rm circ}$ is the circular velocity at a given radius, 
and $E_{\rm circ}$ is the energy of a star in a circular orbit at $(R,z)=(R_{\odot},0)$. 
\autoref{fig:gaia_quant_eorb}(a,b) show the density profiles. At least 8 peaks can be identified and these 
are marked with vertical dotted lines. The peak at $E'=-0.156$ corresponds to the Hercules stream and is shown with a different color. \autoref{fig:gaia_quant_eorb}(c) shows the profile of
mean vertical velocity. A large-scale trend of increase in $V_z$ with $E$  can be seen similar to \citet{2018MNRAS.478.3809S} who studied $V_z$ as a function of  $L$. Note, for a given $R$ and $V_{R}$, $E$ increases monotonically with $L$, and here 
the range of $R$ is almost constant and $V_{R}$ is small. This large-scale trend of $V_z$ is due to the warp.

Besides the large-scale trend, peaks at $E'=[-0.021,0.097]$ can also be seen. The location of these peaks matches with peaks seen in density. 
\autoref{fig:gaia_quant_eorb}(d) plots median 
vertical distance z. There is no large-scale trend, but 
3 peaks ($E'=[-0.21,0.097,0.189]$) are clearly identifiable and they match with the peaks in density. Two of the peaks also match with peaks in $V_z$. \autoref{fig:gaia_quant_eorb}(e) shows the median value of $-|z|$. Almost all density peaks have a corresponding peak in this plot, which is a reflection of the fact that the stars in the density peaks lie close to the Galactic plane. Finally, \autoref{fig:gaia_quant_eorb}(f) shows the profile of median $V_{R}$. Although all peaks do not match in location with all peaks in density, however, for each undulation in the profile of density there is an undulation in $V_{R}$. {\it This indicates that $V_{R}$ and the density peaks are strongly correlated with each other.}
Note, a consequence of $V_{R}$ peaks not matching up with density peaks is that stars in a ridge are not symmetrically distributed about $V_R$, and arches in the
\vphivR{} plane show such a behaviour.

\subsubsection{Dissection in elemental abundances with GALAH}
\label{sec:ridge_chem}

We now study the elemental abundance in the \vphiR{} plane. In \autoref{fig:vphiR_galah}a, \vphiR{} is mapped by [Fe/H]. For the region 200$<$\vphi/\kms{}$<$250 the background metallicity is around [Fe/H]$\approx-0.1$, reflecting the local ISM around the solar neighbourhood which is sub-solar \citep{2012A&A...539A.143N}. The ridges in this region however, are mainly composed of solar metallicity  stars, with typical [Fe/H]$\approx$0.03. In \autoref{fig:vphiR_galah}b \vphiR{} is mapped by [$\alpha$/Fe]. The ridges stand out as a population with [$\alpha$/Fe]$\approx0.05$ (close to solar values). This is consistent with the ridges being made of stars that lie predominantly in the plane. Stars close to the plane are younger, and young stars are metal rich and alpha-poor \citep[age-scaleheight and age-metallicity relations, e.g.,][]{2017MNRAS.471.3057M}. 

Beyond \angz{}$= 2080$ kpc \kms{} there is a sharp cut-off in metallicity (black dotted curve, \autoref{fig:vphiR_galah}a). This region is dominated by relatively metal-poor stars with typical [Fe/H] $\approx-0.3$, and is also alpha-enhanced around [$\alpha$/Fe]$\approx0.1$ (\autoref{fig:vphiR_galah}b). This suggests that the origin of these stars is different from those along the ridges. \angz{}$>2080$ kpc \kms{} corresponds to a guiding radius  $R_{\rm G}>9.5$ kpc (assuming a flat rotation curve). These stars thus belong to the outer disc and their low metallicity is consistent with the Galaxy's negative metallicity gradient with $R$ \citep{2014AJ....147..116H}.

Similarly, stars at the bottom of \autoref{fig:vphiR_galah}(a,b) with \vphi{}$<$ 150 km s$^{-1}$ also show a sharp change in abundances. These stars have large asymmetric drift, are rotating slowly, and have [$\alpha$/Fe] $>$ 0.14 and [Fe/H] $< -0.4$. These properties  are consistent with that of the traditional thick disc, which is metal-poor, alpha-enhanced, and kinematically hot \citep[][]{2014A&A...562A..71B,2018MNRAS.476.5216D}. 

\graphicspath{{figures/}} 
\begin{figure}
\includegraphics[width=1.0\columnwidth]{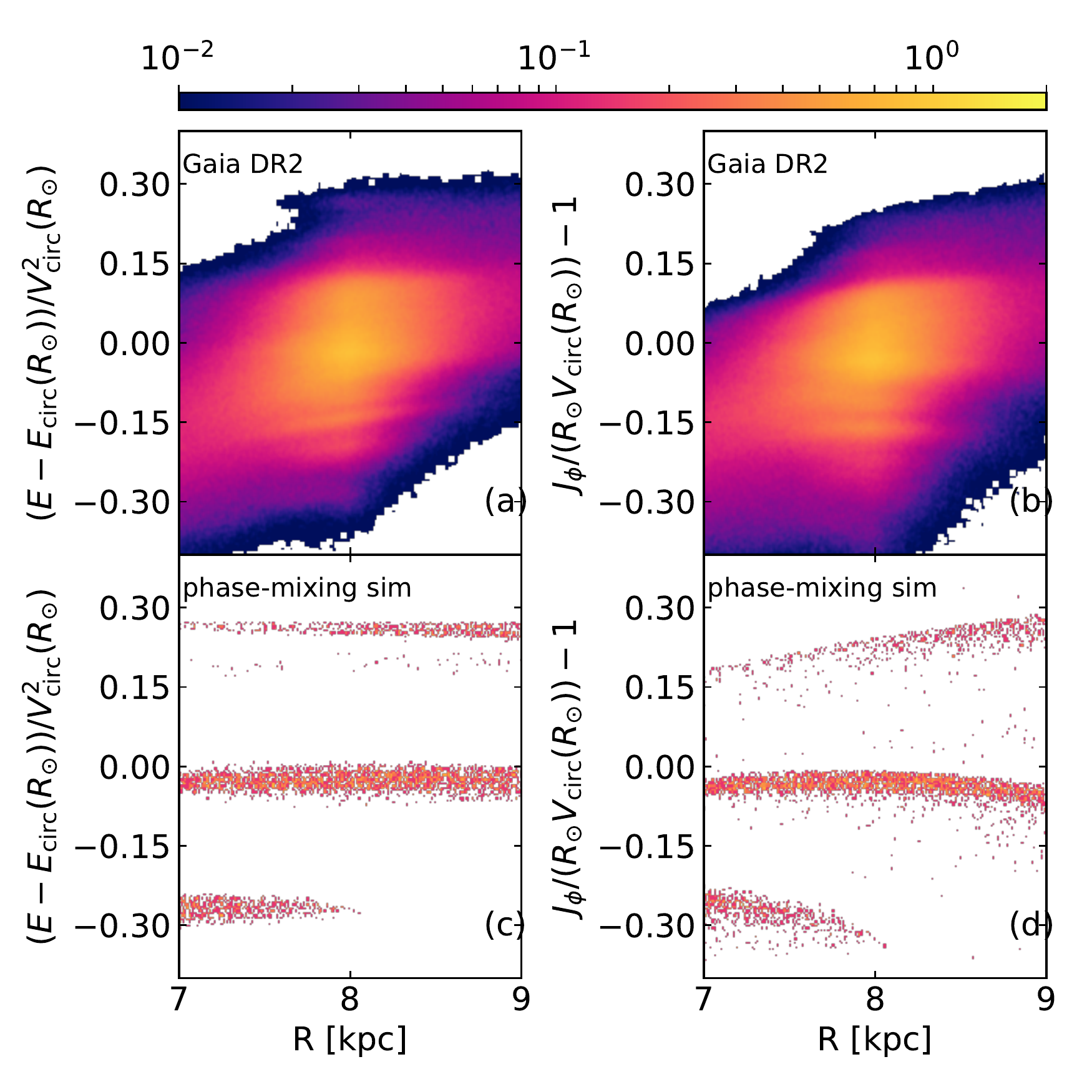}
\caption{ Distribution of stars in $(R,E)$ 
and $(R,J_{\phi})$ space for; a) \gaia{}, and b) phase mixing simulation of a single spiral arm after an  evolution of $\tau=650$ Myr. In \gaia{} data some ridges have constant angular momentum while some have constant energy (specially ridges at large $|E|$). In contrast, phase mixing generates ridges that have constant energy.\label{fig:gaia_er_lr}}
\end{figure}

\subsection{Analysis of the \vphiR{} plane using a phase-mixing simulation}
\label{sec:toy_model_results}
We now consider a toy model of phase mixing similar to that used by \citet{2018Natur.561..360A} to explain some of the features seen in the \vphiR{} plane. We consider an initial distribution of particles confined to four thin spiral arms;
in \citet{2018Natur.561..360A} the particles were confined to a single line. The  particles are then evolved in time under a multi-component analytic potential. The simulation is designed to  mimic phase mixing of perturbations caused by transient spiral arms (for further details see \autoref{sec:toy_model}). 

The distribution of stars in the \xy{} and the \vphiR{} planes are shown in \autoref{fig:phasemix_ridges}
for four different snapshots in time. Also shown are maps of $\langle V_R \rangle$ in  the \vphiR{} plane; see \autoref{fig:phasemix_ridges}(i-l). As we move forward in time, the spiral pattern decays (\autoref{fig:phasemix_ridges}(a-d)), and the ridges start to form and they increase in number and become more stretched and therefore thinner (\autoref{fig:phasemix_ridges}(e-h)). The ridges can also be seen in maps of $\langle V_R\rangle$. The ridges are approximately linear in the \vphiR{} plane and resemble lines of constant angular momentum. The appearance of the ridge structure is a consequence of phase mixing and can be understood in terms of \Liouv{}'s theorem, which states that the full phase-space density (or volume) of a system evolving in a fixed potential is conserved. In the case of our simulation, the phase space is made of $(X,Y, V_X, V_Y)$. Initially the density in the \xy{} space is high while that in \vphivR{} space is low. As the spiral pattern disperses, the density in the \xy{} plane  reduces, but to conserve the phase-space density, the density should increase in other dimensions. The structures in the \vphiR{} are in some sense a reflection of this phenomenon. 

\autoref{fig:phasemix_ridges}(m-p) show the distribution of stars in orbital energy and Galactocentric radius $(E,R)$ plane. Discrete energy levels can be seen. Stars in a ridge lie in a narrow energy interval; we explore this further in \autoref{fig:gaia_er_lr}. The figure shows the distribution of stars in the $(E,R)$ and $(J_{\phi},R)$ plane. 
The top panels show results for \gaia{}, while the bottom panels are for the phase-mixing simulation where
we only show stars belonging to a single spiral. It is clear from \autoref{fig:gaia_er_lr}(c,d) that in phase-mixing simulations, the ridges are curves of constant energy rather than constant angular momentum.  According to \autoref{fig:gaia_er_lr}(c,d), near $E'=0$, constant energy and constant angular momentum curves are both expected to be flat, it is only 
at higher values of $|E'|$ that one can differentiate between the two cases. 
For the observed data, ridges between $-0.15<E'<0$ (corresponding to the Hercules stream) look flat in $J_{\phi}$, while the rest of the ridges, especially with large values of $|E|'$, are flatter in $E'$ than in $J_{\phi}$. 
This suggests that different ridges can 
originate from different physical processes. The lower most ridge (vertical coordinate of -0.3) in \autoref{fig:gaia_er_lr}b is slanted downwards in $J_{\phi}$ just like in \autoref{fig:gaia_er_lr}d, while the topmost ridges in \autoref{fig:gaia_er_lr}b is slanted outwards just like in \autoref{fig:gaia_er_lr}d. 
For the lower two ridges with $E'<-0.2$, it is clear that the ridges are sharper in energy 
than in angular momentum, lending further support to 
the ideas that  energy as  a quantity is better than angular momentum for characterizing these ridges.

\subsection{Analysis of the \vphiR{} plane using disc  N-body simulations}
\label{sec:nbody_results}

\graphicspath{{figures/}} 
\begin{figure*}  
\includegraphics[width=2.0\columnwidth]{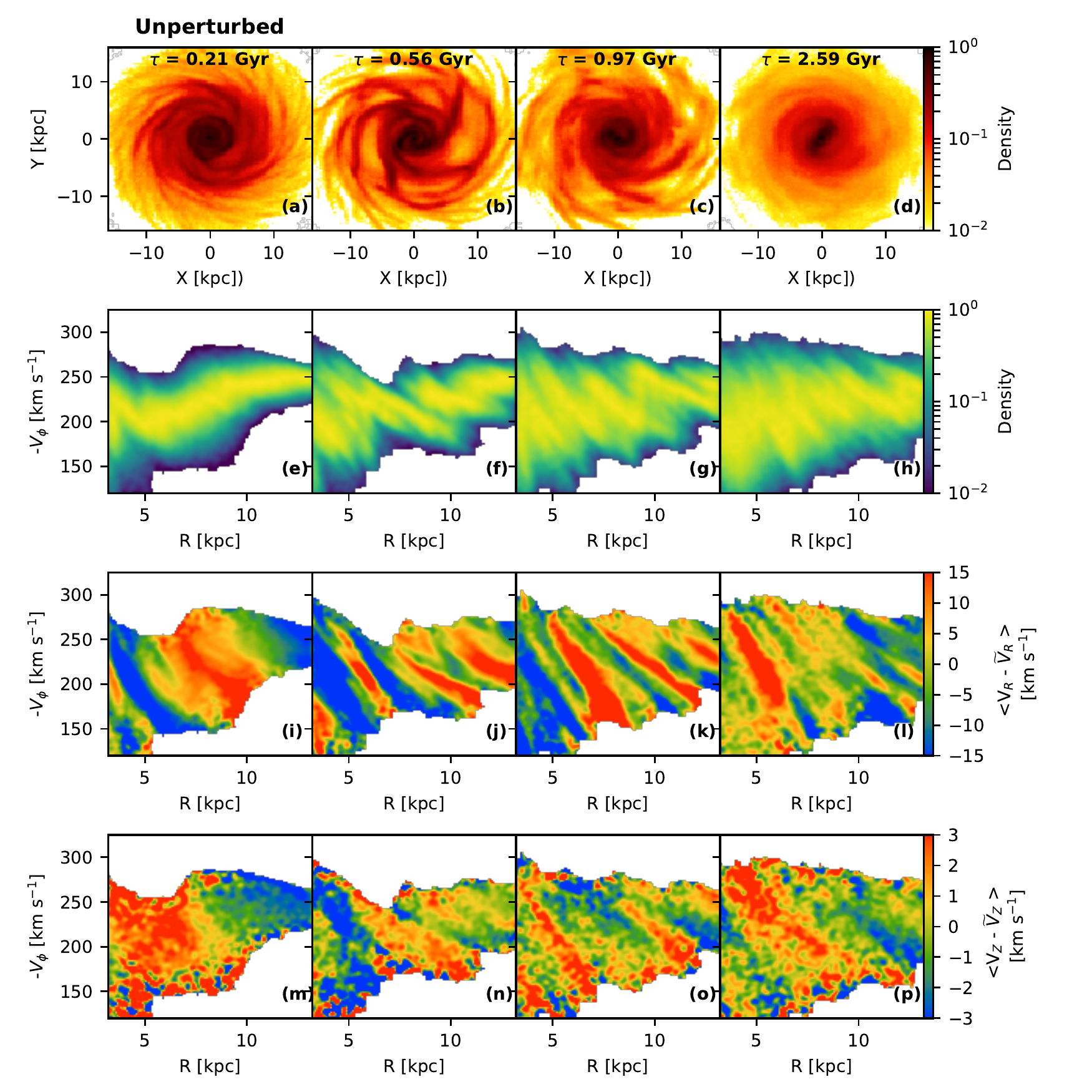}
\caption{ 
Evolution of stars in the N-body simulation of an unperturbed disc (Model P). Panels(a,b,c,d) show the density in $xy$ plane for four different snapshots at $\tau=[0.21,0.56,0.97,2.59]$ Gyr. Panels(e,f,g,h) show $p(V_{\phi}|R$), the 
probability density of $V_{\phi}$ conditional on $R$. Panels(i,j,k,l) show the \vphiR{} plane mapped by \mnvR{}. Panels(m,n,o,p) show the \vphivR{} plane mapped by \mnvz{}. In Panels(e,f,g,h,i,j,k,l,m,n,o,p), stars are selected in the region $(|R-R_{\odot}|{\rm /kpc} <5.0)\&(|\phi-\phi_{\odot}|<25 ^{\circ})$. 
\label{fig:iso_nb_ridge}}
\end{figure*}

\graphicspath{{figures/}} 
\begin{figure*}  
\includegraphics[width=2.0\columnwidth]{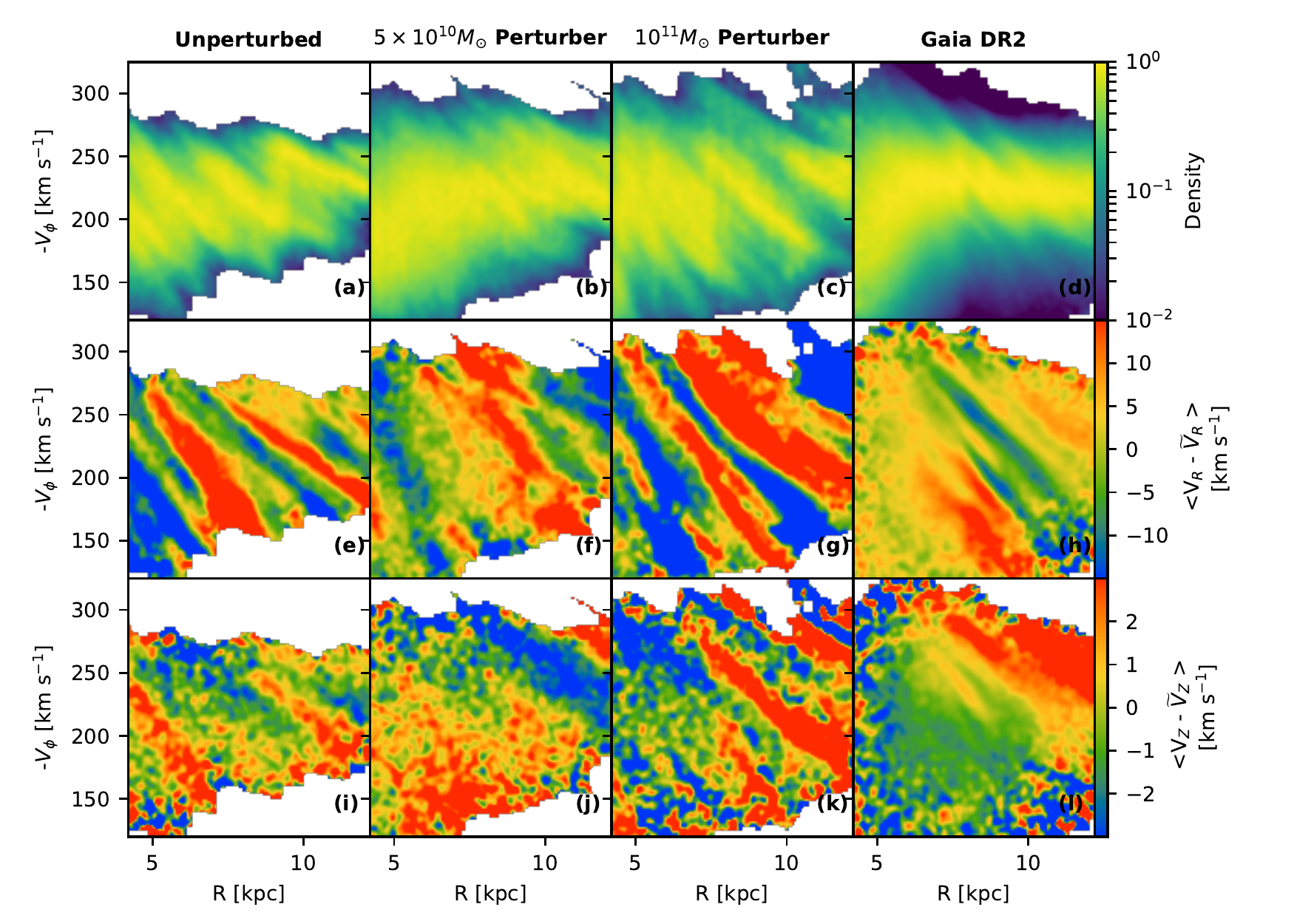}
\caption{Comparison of features in the \vphiR{} plane seen in N-body simulations with the observed data. Snapshots shown in panels(a,b,c) are at $\tau=0.97$ Gyr, $\tau=1.53$ Gyr, and $\tau=1.54$ Gyr, respectively. In all panels, stars are selected in the region $(|R-R_{\odot}|{\rm /kpc}<4.0)\&(|\phi-\phi_{\odot}|<25 ^{\circ})$. 
Panels (a,b,c,d) show $p(V_{\phi}|R$), the probability density of $V_{\phi}$ conditional on $R$.
\label{fig:multi_nb_ridge_compare}}
\end{figure*}

\graphicspath{{figures/}} 
\begin{figure*}  
\includegraphics[width=2.0\columnwidth]{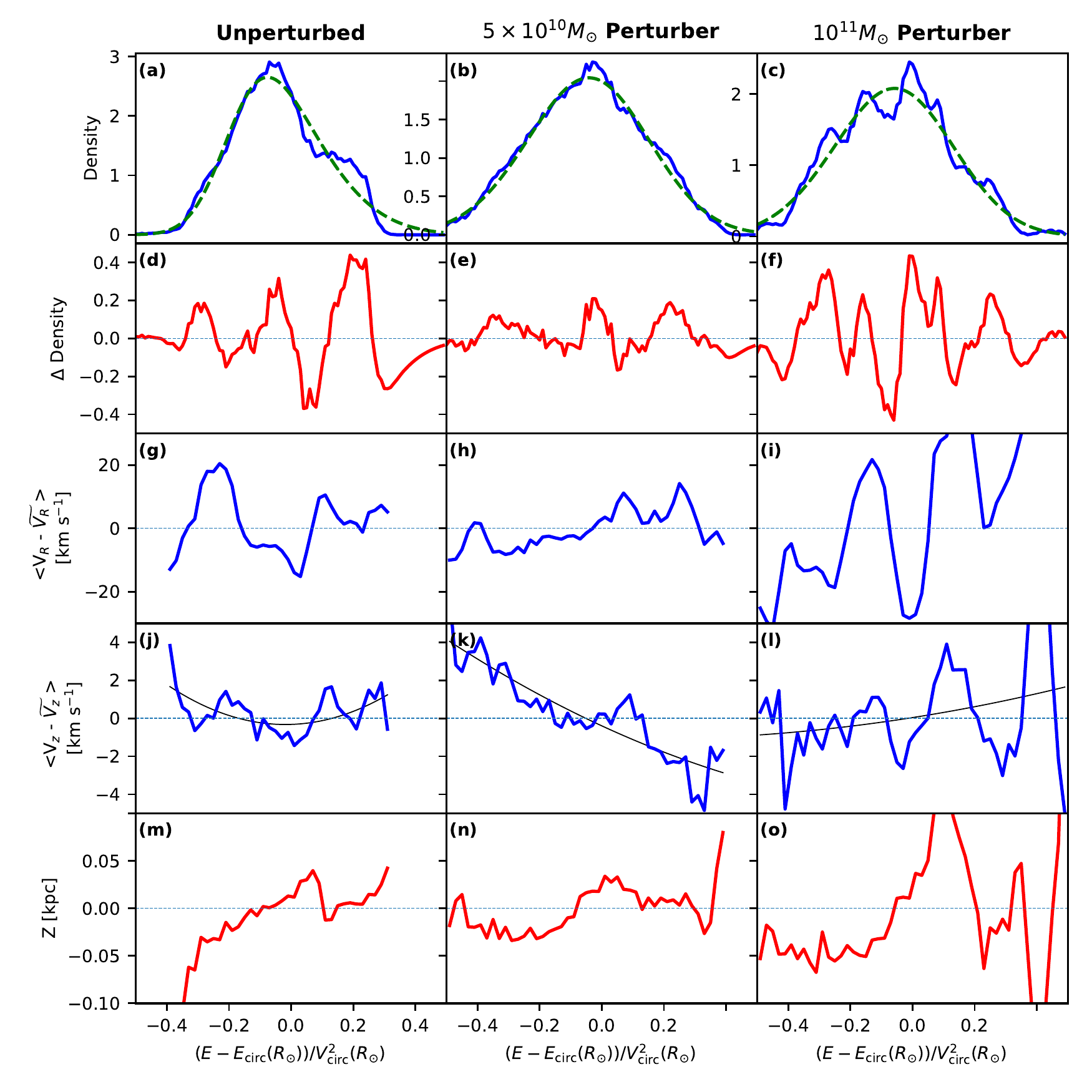}
\caption{Profiles of different quantities as a function of orbital energy for stars in selected from disc N-body simulations. Results for galaxy that is unperturbed (Model P, left column panels), perturbed by inetermediate mass Sgr (Model S, middle column panels) and high mass Sgr (Model R, right column panels) are shown. Snapshots shown in panels(a,b,c) are at $\tau=0.97$ Gyr, $\tau=1.53$ Gyr, and $\tau=1.54$ Gyr, respectively. Panels(a-c) show the Energy distribution with a best fit skewed normal distribution (green dotted line). Panels(d,e,f) show the residuals in density distribution after subtracting the skewed normal. Panels(g-i) show the median vertical velocity, with a background fit for warp. Panels(j,k,l) show the vertical distance from the plane. Panels(m,n,o) show the median radial velocity profile. The figure is analogous to \autoref{fig:gaia_quant_eorb}.
\label{fig:dens_correl}}
\end{figure*}

We now consider the more realistic N-body simulations of the Galaxy described in \autoref{sec:nbody_model}. The first scenario, Model P, is that of an isolated galaxy, i.e., unperturbed by a satellite. The \xy{} density of four selected snapshots at $\tau=[0.21,0.56,0.97,2.59]$ Gyr are shown in \autoref{fig:iso_nb_ridge}(a-d). At $\tau=0.21$ Gyr (\autoref{fig:iso_nb_ridge}a), the disc settles into an equilibrium configuration and develops tightly wound spiral arms. Such self-excited instabilities forming spiral arms are a known feature of N-body simulations in disc galaxies \citep{2012ApJ...751...44S}. The corresponding \vphiR{} density map (\autoref{fig:iso_nb_ridge}e) is largely uniform and lacks ridge-like substructure as seen in \gaia{} (in \autoref{fig:vphiR_gaia}). Similarly, in the velocity maps (\autoref{fig:iso_nb_ridge}(i,m)), there are fine-structure blobs in the kinematics with \mnvR{} $\approx 10$ \kms{} and \mnvz{} $\approx 3$ \kms{}, but no ridges can be seen. 

By $\tau=0.56$ Gyr, the spiral arms have weakened slightly, they are fewer and thicker (\autoref{fig:iso_nb_ridge}b). Interestingly, the \vphiR{} density at this snapshot shows large-scale diagonal stratification, with a span of about 4 kpc (\autoref{fig:iso_nb_ridge}f). The \mnvR{} map shows multiple thin diagonal ridges with an alternating pattern of radially outward and inward motion (\autoref{fig:iso_nb_ridge}(j)).

By the next snapshot at $\tau=0.97$ Gyr, the spiral arms are found to have diffused and weakened  (\autoref{fig:iso_nb_ridge}c). The corresponding \vphiR{} density map shows several prominent ridges that extend over $5<R/kpc<15$ and have a more linear appearance compared to the previous snapshot (\autoref{fig:iso_nb_ridge}g).  The ridges are also clearly present in the \mnvR{} and \mnvz{} maps, where the amplitude of the radial oscillations is again higher than  the vertical component.

By the final snapshot, chosen at $\tau=2.59$ Gyr, the spiral arms are found to have significantly decayed. A central bar with half-length of $\sim 2.5$ kpc is visible prominently (\autoref{fig:iso_nb_ridge}d). The density, \mnvR{}, and \mnvz{} maps continue to show large scale ridges (\autoref{fig:iso_nb_ridge}(h,i,p)). In summary, \autoref{fig:iso_nb_ridge} shows that an unperturbed galaxy can reproduce ridges in the \vphiR{} plane with features similar to that seen in \gaia{}. 
The ridges appear as the spiral structure decays, and are maintained as long as this decay is going on. As was already mentioned in \autoref{sec:toy_model_results}, this is a consequence of \Liouv{}'s theorem which requires that the density in phase-space is always conserved. This suggests that internal instabilities such as transient spiral arms, could be responsible for the ridges seen in \autoref{fig:vphiR_gaia}.

Next, we consider the scenario where the Galaxy is tidally perturbed by an orbiting satellite. Model S simulates the interaction with an intermediate mass Sgr galaxy ($5 \times 10^{10} M_{\odot}$), while Model R simulates the interaction with a heavier Sgr  galaxy ($10^{11} M_{\odot}$). In both cases, Sgr crosses the disc at around $\tau=0.15$ Gyr, and perturbs the galaxy from its equilibrium state. Previously, in simulations run in \cite{2019MNRAS.tmp..222B}, we noted that disc crossing by Sgr wipes out previous coherent structure and generates new structures in the Galaxy. Evolving the galaxy for $\tau=1.5$ Gyr, allows for enough time to develop, decay, and phase mix the spiral arms as well as the effects of Sgr. 
For this reason we compare the unperturbed and perturbed scenarios at roughly coeval timestamps of ($\tau=1.5$ Gyr), i.e., allowing for enough time for perturbations to phase mix. 

\autoref{fig:multi_nb_ridge_compare} shows the density, \mnvR{}, and \mnvz{} maps in the \vphiR{} plane for the various N-body simulations alongside \gaia{} data. We note the presence of ridges in all three simulations (\autoref{fig:multi_nb_ridge_compare}(a,b,c)). Ridges are also present in the maps of \mnvR{} (\autoref{fig:multi_nb_ridge_compare}(f,g)) and \mnvz{} (\autoref{fig:multi_nb_ridge_compare}(j,k)).
The ridges for \gaia{} data in (\autoref{fig:multi_nb_ridge_compare}d appear to be smeared out at the edges. This is due to observational errors in proper motion and parallaxes that are dominant at larger distances. 
We would like to point out, that the purpose of the N-body simulations in this paper, is to demonstrate what kinematic signatures can be generated at a given location in the Galaxy. These are not, however, selection function matched snapshots i.e., the particles do not have stellar parameters/magnitudes assigned that we could convolve \gaia{}-like errors. In any case, our aim her is not to match the exact number of features or their location one-to-one, which would be affected by the smear due to observational errors.

In \autoref{fig:gaia_quant_eorb}, we saw that for \gaia{}, ridges are correlated in kinematics and spatial density. In \autoref{fig:dens_correl} we explore similar correlations for our N-body simulations. We select stars around $(R,z,\phi)=(8.2,0.0,180.0^{\circ})$ and consider the profiles of \mnvz{}, \mnvR{}, and $z$ against the dimensionless orbital energy, $E'=$\energy{}. For all simulations, peaks can be seen in profiles of density, \mnvz{}, $z$, and $V_{R}$.
It is worth noting that the unperturbed model has no tidal interactions, hence, the observed vertical oscillations for the unperturbed model must be due to internal processes. 

A number of features seen in \autoref{fig:gaia_quant_eorb} for the \gaia{} 
data can also be seen in the simulations. The location of peaks in $z$ match with location of peaks in $v_{z}$. 
Location of extrema in $v_{z}$ match with location of peaks in density. For the the unperturbed case it is the minima that matches and for the high mass case it  is the maxima. For the intermediate mass case we do not see such an association. 
We note that the matching of the location of peaks in $z$ and $v_{z}$ is not a general feature, because it was only seen at a few special locations within the galaxy. 

The \mnvz{} profiles (\autoref{fig:dens_correl}(j-l)) show a large scale trend like in the \gaia{} data. Such a trend is expected for the presence of a warp. A clear warp was detected in all our simulations. A plot of mean $z$ and $v_{z}$ as a function of $\phi$ showed a sinusoidal pattern with the $v_{z}$ profile being shifted by $90^{\circ}$ with respect to the $z$ 
profile. 

The amplitude of fluctuations for all the plotted quantities (density, \mnvz{}, $z$, and $V_R$), 
is considerably higher for the high mass Sgr case compared to the other two simulations. A comparison with \autoref{fig:gaia_quant_eorb} 
shows that the amplitude of \mnvz{}, $z$, and $\langle V_R\rangle$ fluctuations for the case 
of \gaia{} is comparable to the case of unperturbed and intermediate mass Sgr simulations, making the case for the high-mass perturber unfavourable.

\graphicspath{{figures/}} 
\begin{figure*}
\includegraphics[width=2.0\columnwidth]{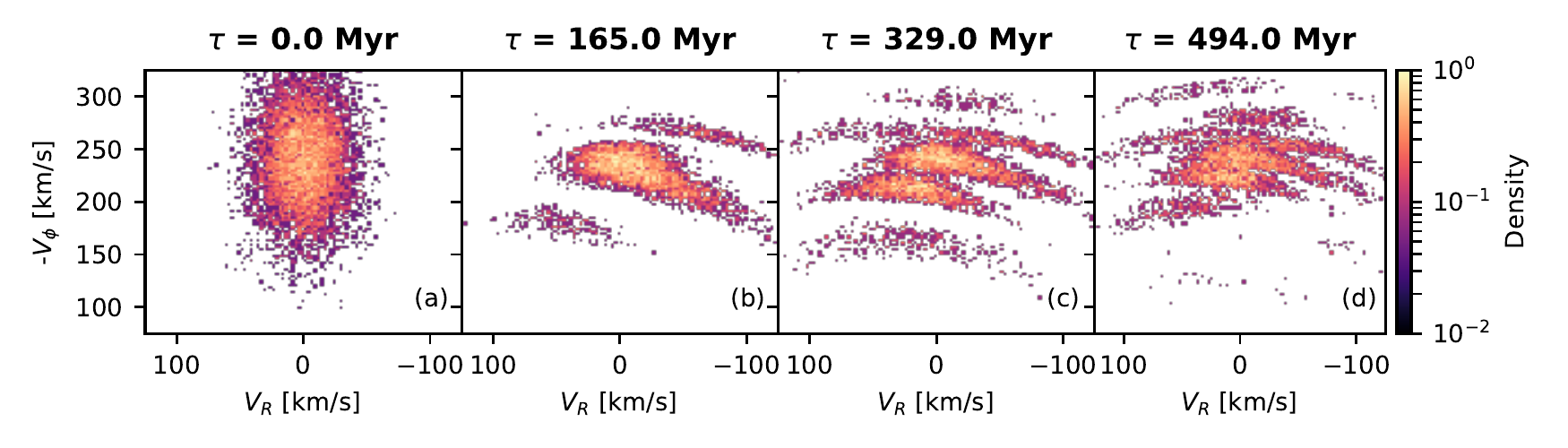}
\caption{Evolution of stars in the phase mixing simulation (\autoref{sec:toy_model}), same as \autoref{fig:phasemix_ridges}, but here we show the \vphivR{} plane. Stars are selected to lie in $(|\phi-180.0^{\circ}|<25^{\circ})\&(|R-8.2|/{\rm kpc}<0.25)$. \label{fig:phasemix_arches}}
\end{figure*}

\subsection{Analysis of the  \vphivR{} plane: arches}

We now study the \vphivR{} plane. 
\autoref{fig:ridge_arch_gaia}b shows the distribution of \gaia{} stars. Arch-like structures can be seen and they are 
asymmetrical about the $V_R=0$.
In \autoref{fig:phasemix_arches}, we show the distribution 
of stars in the phase-mixing simulation. Initially, there are no arches, but as time proceeds, arches start to appear, increase in number, and become thinner. In the final snapshot, at 494 Myr, multiple arches are clearly visible, and they also appear to be asymmetrical as in the observed data.

\begin{figure*}  
\includegraphics[width=2.0\columnwidth]{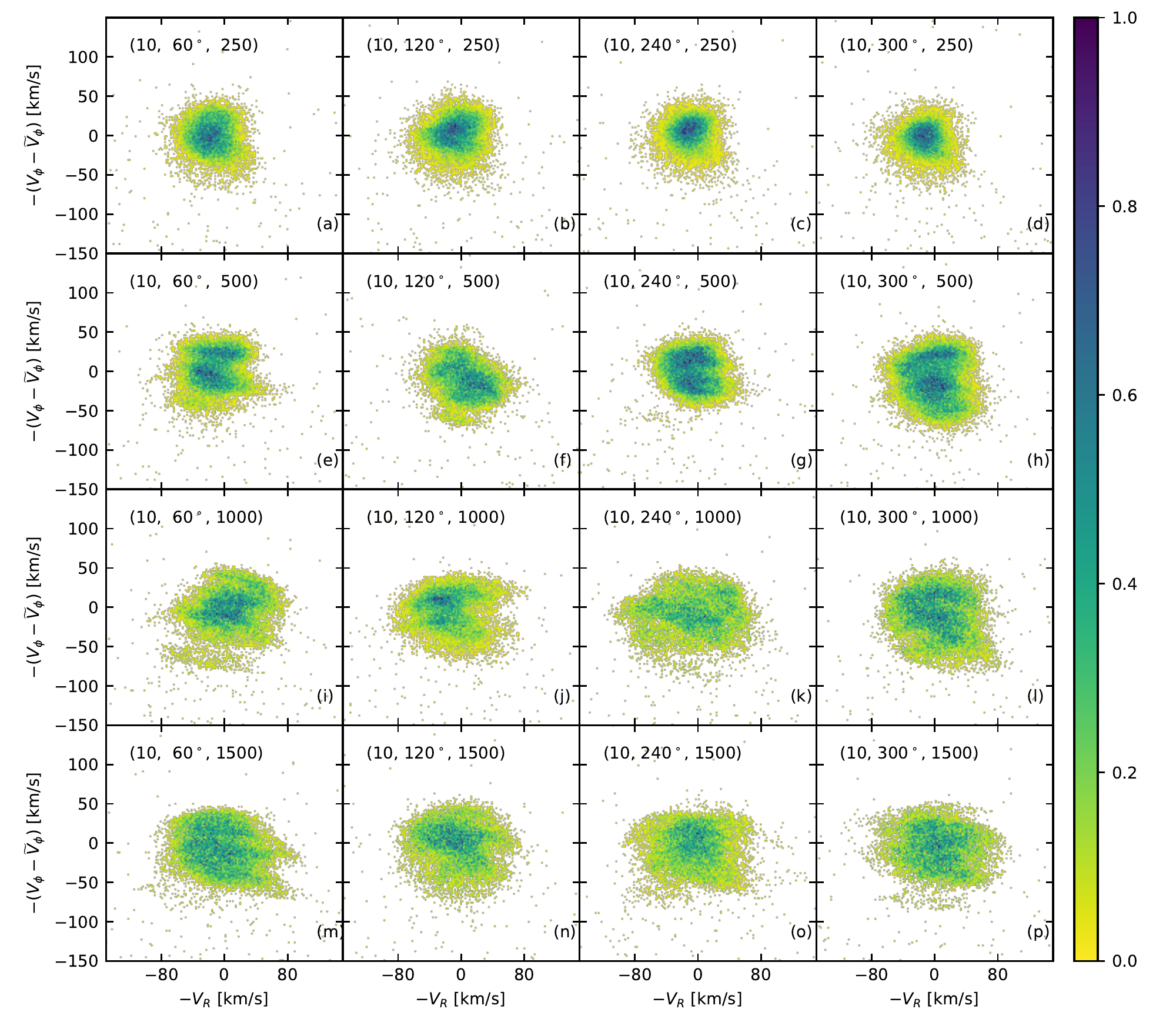}
\caption{Distribution of stars in the \vphivR{} plane for the N-body simulation P (Unperturbed Galaxy). 
Distributions for four different times and at various different azimuth angles are shown. The Galactocentric radius $R$, the azimuth angle $\phi$ and the time in Myr are labelled on the plots.
Stars were restricted to $(|\Delta R|<0.25 {\rm kpc})\&(|\Delta \phi|<25^{\circ})$.  \label{fig:arch_mwonly}}
\end{figure*}

We now study the distribution of stars in the \vphivR{} space using disc N-body simulations. \autoref{fig:arch_mwonly} shows the distribution of stars for simulation P.
We show snapshots corresponding to time $\tau$ of 250, 500, 1000 and 1500 Myr. 
For each time, we show distributions at four different locations in azimuth. 
The simulation starts with a smooth disc and by 250 Myr strong tightly wound spiral arms can be seen, however the velocity distribution is devoid of any substructures at this stage. As the simulation evolves, the velocity distribution becomes irregular and develops substructures. Arches are visible in all snapshots with $\tau>=500$ Myr and they are not symmetric about the $V_R=0$.

\begin{figure*}  
\includegraphics[width=2.0\columnwidth]{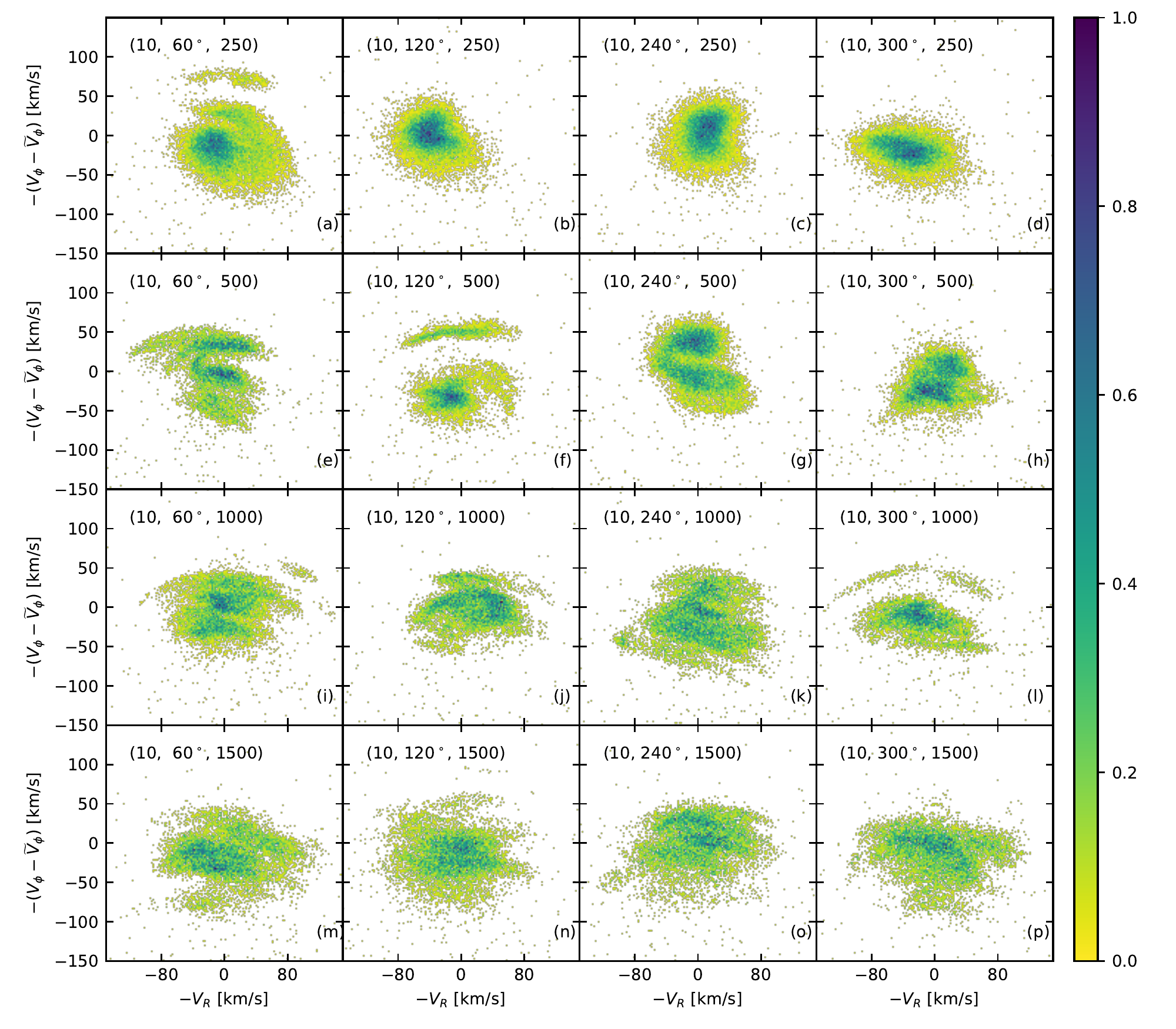}
\caption{Analogue of \autoref{fig:arch_mwonly} but for stars in the N-body simulation S (interaction with an intermediate mass satellite). \label{fig:arch_s}}
\end{figure*}

\begin{figure*}  
\includegraphics[width=2.0\columnwidth]{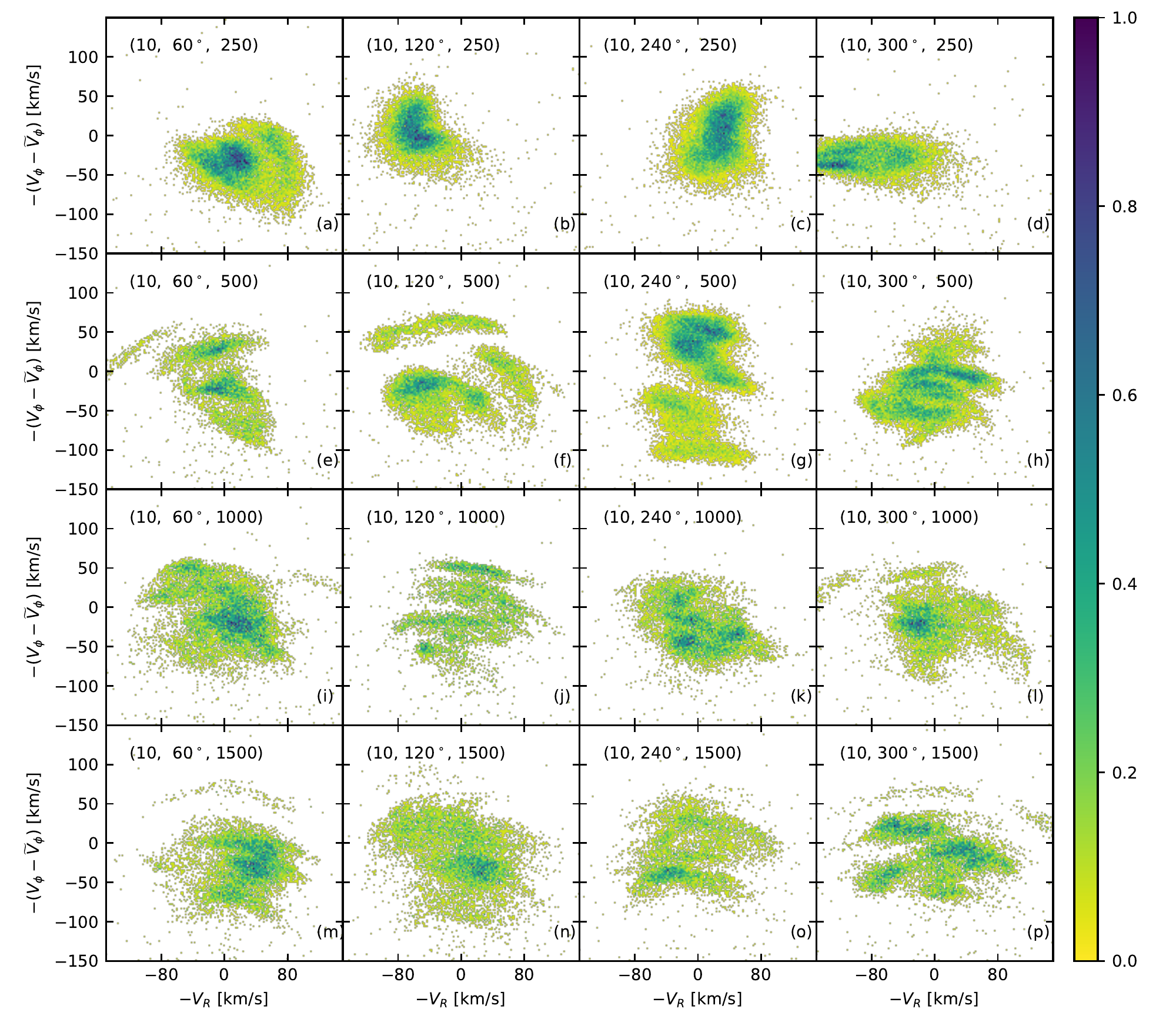}
\caption{Analogue of \autoref{fig:arch_mwonly} but for  stars in the N-body simulation R (interaction with a high mass satellite).\label{fig:arch_r}}
\end{figure*}
\autoref{fig:arch_s} shows the distribution of stars for simulation S that corresponds to interaction with an intermediate mass satellite. Similarly, \autoref{fig:arch_r} shows the distribution for simulation R 
that corresponds to interaction with a high mass satellite. As compared to the simulation P of the unperturbed galaxy,  considerably more substructures and arches can be seen 
in simulations S and R. The simulation R with high mass shows more arches  
than simulation S. It is clear that arches can develop even when there is no external perturber, but when a perturber like an orbiting satellite is present, the arches are stronger and more numerous. The fact that the number and strength of arches depends upon the mass of the satellite means that we can use the observed data to put limits  
on the satellite mass. From our set of simulations, we conclude that simulation P has too few arches and simulation R too many, and it is the simulation S that matches best with the \gaia{} data.

\subsection{Analysis of the $(R,V_R)$ plane}
\begin{figure*}  
\includegraphics[width=2.0\columnwidth]{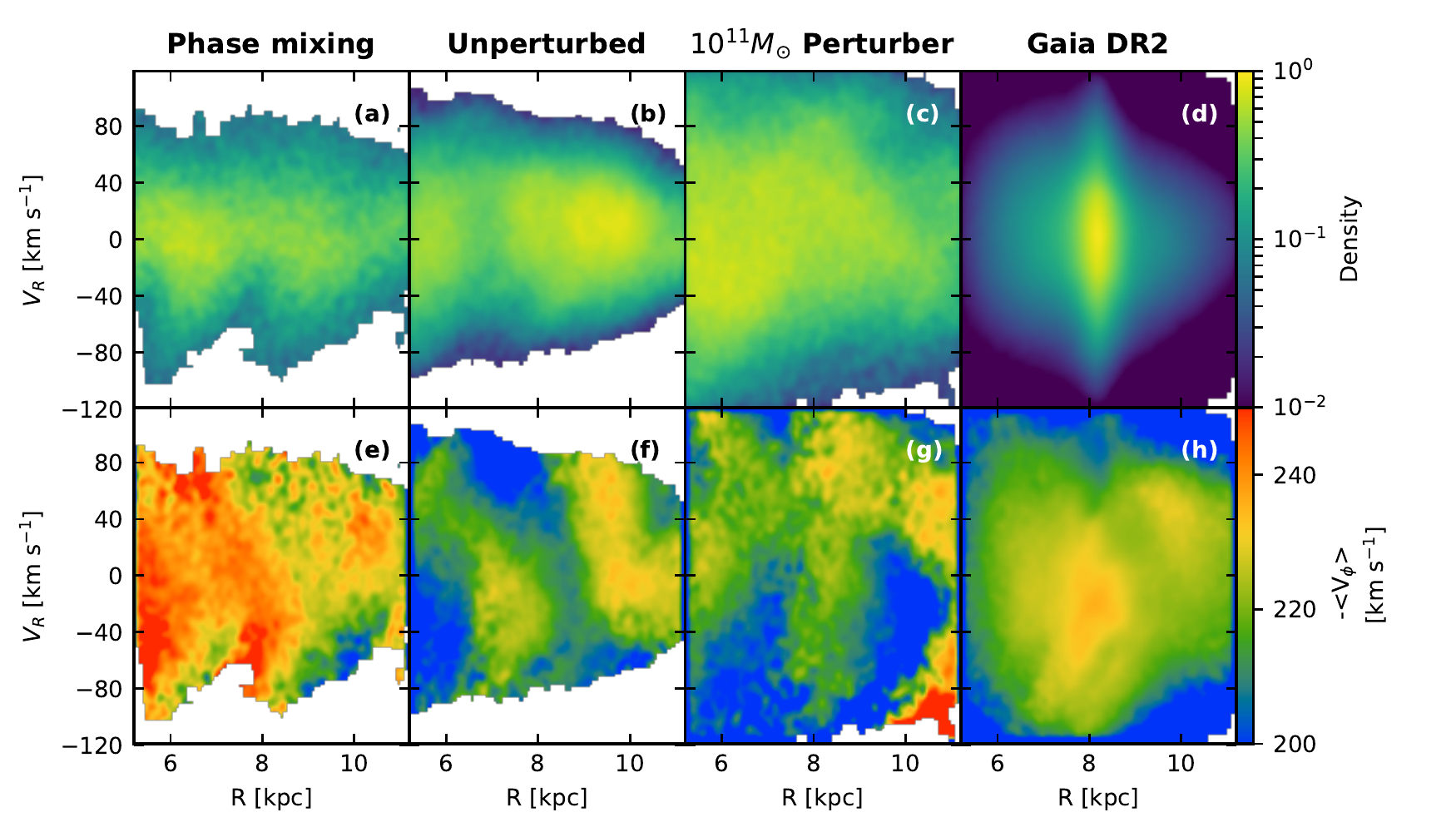}
\caption{Maps of density and $\langle V_{\phi} \rangle$ in the $(R,V_R)$ space. (a,e) Phase mixing simulation ($\tau=494$ Myr). (b,f) Simulation of an unperturbed disc ($\tau=0.97$ Gyr). (c,g) Simulation of a disc perturbed by a high mass satellite ($\tau=1.54$ Gyr) (d,h) Gaia DR2.  The density maps are smooth, but the $\langle V_{\phi} \rangle$ maps show cone like structures (similar to a rotated pine tree), which are also predicted by phase mixing simulations.
 \label{fig:rvr}}
\end{figure*}

Structures have been reported in the \vphiR{} and 
$(z,V_z)$ planes, but so far the $(R,V_R)$ space has not been explored. \autoref{fig:rvr} shows maps of density 
and $\langle V_{\phi} \rangle$ in the $(R,V_R)$ phase space. The density map is extremely smooth and shows no substructure in \autoref{fig:rvr}(a-d). However, arrow shaped substructures can be seen in the $V_{\phi}$ map in \autoref{fig:rvr}(e-h). Phase mixing can explain the substructures seen in this space. Such substructures can also be seen in disc N-body simulations. These substructures provide additional independent constraints on models trying to explain the origin of phase-space substructures in the Galaxy.

\section{Discussion and Conclusions}
\label{sec:discussion}
We have explored the ridge-like features in the
\vphiR{} plane using position and velocities from \gaia\ and elemental abundances from \glh{}. We find that ridge-like features are visible not only in the density maps but also in maps of $\langle V_R \rangle$ , $\langle |z| \rangle$, $\langle V_z \rangle$, [Fe/H] and $[\alpha/{\rm Fe}]$ (\autoref{fig:vphiR_gaia} and \autoref{fig:vphiR_galah}). Ridges in the $\langle V_R \rangle$ map are more prominent and visible to much larger Galactocentric radii $R$ than in the density map. The $\langle |z| \rangle$ map suggests that the ridges are more prominent for stars close to the mid-plane of the Galaxy. The \glh{} data suggest that stars in the ridges are predominantly of higher metallicity than the non-ridge stars ($\sim$solar [Fe/H]) and solar $[\alpha/{\rm Fe}]$ (\autoref{fig:vphiR_galah}). Since, typically stars close to the plane have values of [Fe/H] and $[\alpha/{\rm Fe}]$ that are close to solar, this explains the trends with elemental abundance. 
That the ridge stars are predominantly at low $|z|$ could be due to one or all of the following three reasons: a) The ridges are due to transient perturbations (i.e., spiral arms) that are close to the plane and are disrupting and phase mixing with time;
b) the ridges are due to interaction of stars with perturbations that are close to the plane; 
c) stars close to the plane are kinematically cold 
and it is easier to perturb them.

Our phase-mixing simulation of disrupting spiral arms can explain a wide array of kinematic features in the observed \gaia\ data. They simultaneously reproduce the ridges in the \vphiR{} plane (\autoref{fig:phasemix_ridges}(e-h)), the ridges in 
the $\langle V_R \rangle$ maps (\autoref{fig:phasemix_ridges}(i-l)), and the arches in the \vphivR{} plane (\autoref{fig:phasemix_arches}). They also reproduce the observed asymmetry in the arches.  While a bar perturbation has been shown to generate ridges, {\it the number of ridges generated from a bar alone are too few to match the observed data} \citep{2018Natur.561..360A, 2018MNRAS.481.3794H}. 
Phase mixing generates surfaces of constant energy and this explains the occurrence of both the ridges and the arches.

More realistic N-body simulations of a disc in which spiral arms are naturally generated support the 
results obtained from phase mixing. In these simulations,  the spiral arms grow in strength with time till about 500 Myr, and then start to decay. As the spiral arms decay and get phase mixed, the ridges and arches are found to grow in prominence, a phenomenon that was also seen in the phase-mixing simulation (\autoref{fig:iso_nb_ridge}).
Our N-body simulations show ridges in the $\langle V_z \rangle$ maps as seen in the observed data. Simulations in which the disc is perturbed by the passage of an orbiting satellite  also show features 
similar to the case of an unperturbed disc (\autoref{fig:multi_nb_ridge_compare}). 
However, the ridges are found to be more pronounced, in both the 
$\langle V_R \rangle$ and $\langle V_z \rangle$ maps, when the mass of the orbiting  satellite is higher. This makes the case of a $10^{11} M_{\odot}$ or higher mass perturber unfavourable but a 
perturber with $5\times 10^{10} M_{\odot}$ is still consistent with \gaia{}.

\cite{2018Natur.561..360A} tentatively suggest that arches in the \vphivR{} plane are projections of ridges in the \vphiR{} plane. We note that, while ridges do suggest existence of discrete values of $V_{\phi}$ in the solar neighborhood, they do not necessarily suggest the presence of arches. It is impossible to deduce the distribution of $V_{R}$ from the distribution of stars in the \vphivR{} plane. We have shown that phase-mixing simulations of disrupting spiral arms not only generate ridges but also arches. The physical property unifying the two features is the energy. A ridge in \vphiR{} and an arch in \vphivR{} are both curves of constant energy. 
The phase mixing of disrupting spiral arms 
generates a regular pattern of peaks in the energy distribution of a sample confined to a narrow range in azimuth (\autoref{fig:phasemix_ridges} o,p). 
A curve of constant energy and constant angular momentum 
both appear as a ridge in the \vphiR{} plane. However, out of the above two, only a constant energy curve will manifest itself as an arch in the \vphivR{} plane. 
Note, we observe stars in a narrow range of azimuth, only stars of certain discrete values of angular frequency will end up in the chosen azimuth range at a given time. The fact that we see discrete energy levels suggest that 
the angular frequency is more strongly correlated with energy than with angular momentum.

Two different techniques, our work using phase mixing  and work by \citet{2018MNRAS.481.3794H} using scattering from a perturbation in the potential, both suggest that transient winding spiral arms can explain the multiple ridges and arches seen in \gaia{}.  Interestingly, transience here is through a process of wrapping up rather than fading away in strength with time as generally thought.

The arches seen in \gaia\ are asymmetrical about $V_R=0$ in the \vphiR{} plane . Phase mixing is generally thought to produce symmetric arches \citep{2018MNRAS.480.3132Q}, as was observed by \cite{2009MNRAS.396L..56M} in their phase-mixing simulations. This is because, for stars on an arch, the orbital energy is approximately fixed, and since $E\sim V_{R}^{2} + V_{\phi}^{2}$, the arches are symmetric. However, we show that phase-mixing simulations can generate asymmetrical arches, and that the asymmetry is both intrinsic and apparent.
The slight intrinsic asymmetry is due to the fact that 
an arch has a finite width in energy and the $V_R$ 
changes systematically with energy. This occurs in the
initial stages when phase mixing is incomplete (\autoref{fig:phasemix_arches} b,c).
The apparent asymmetry is due to the following reason and is responsible for asymmetry seen at later stages of phase mixing (\autoref{fig:phasemix_arches} d). The arch due to a single ridge and a single spiral arm is in general symmetrical in the \vphivR{} plane, but the number density of stars is not symmetrical about $V_R=0$. Moreover, the arch  is short and does not span the full range of $V_R$. When multiple arches from different spiral arms are superimposed they look like a large arch with a strong asymmetry.

We also see asymmetrical arches in N-body simulations in which a disc is evolved in a Milky Way like potential, which includes a live dark matter halo (\autoref{fig:arch_mwonly}). Asymmetrical arches were reported by \citet{2011MNRAS.417..762Q} using similar simulations, but they did not study the effect of an interaction with a satellite. \citet{2019MNRAS.485.3134L} studied simulations with an orbiting satellite and reported the presence of ridges but found very few clear arches. We studied simulations 
both with and without an orbiting satellite. We found that simulations in which the disc is perturbed by an orbiting satellite generates more arches. A high mass satellite generates more arches (\autoref{fig:arch_r}) than a satellite with lower mass (\autoref{fig:arch_s}). A  $5 \times 10^{10} M_{\odot}$ satellite was found to describe the observed data the best. Arches develop within 250 Myr of interaction with a satellite, and are clearly visible even after 1 Gyr. 
\citet{2018MNRAS.481.3794H}, using backward integration 
of test particles in a winding spiral arm potential \citep{2000AJ....119..800D}, also reach a similar conclusion.

\cite{2018Natur.561..360A} used the $V_{\phi}$ separation of consecutive ridges and  \cite{2009MNRAS.396L..56M} used the $V_{\phi}$ separation of arches to conclude 
that the perturbation must be older than 1 Gyr and most likely about 2 Gyr. These conclusions are based on the assumption that the ridges are generated by a single perturber. If the ridges and arches are caused by more than one transient spiral arms, then each arm will have its own set of ridges and the separation between the ridges can be smaller as compared to the case of a single perturber for any given age of the perturber. Hence, the $V_{\phi}$ separation cannot be used to reliably date the perturber.

One of the most interesting results of our study is the existence of ridges in the $\langle V_z \rangle(R,V_{\phi})$ maps (\autoref{fig:vphiR_gaia}c). At a given $R$ when $\langle V_z \rangle$ is plotted as function of angular momentum or energy the ridges show up as undulations with clearly defined peaks and valleys (\autoref{fig:gaia_quant_eorb}).
In addition to undulations, a smooth large scale trend is also seen, the $V_z$ increases with $L$ for $L/(V_{\rm circ}(R_{\odot})R_{\odot}) > 1$. This rise  of $V_z$ has been associated with the onset of a warp \citep{2017A&A...601A.115P,2018MNRAS.478.3809S,2018MNRAS.481L..21P}. However, the origin of the undulations is not clear. The data shows that the locations of at least two and possibly three $V_z$ peaks  coincide with the density peaks. This can be interpreted as ridges having a net upward motion. Undulations are also seen in profiles of $z$ with energy. Three peaks are clearly identifiable in $z$ and they match with peaks in $V_z$. Such a coupling of peaks between $V_z$, $z$ and $V_{\phi}$, is also observed in our N-body simulations, of both the unperturbed and the perturbed disc, but infrequently. We could see such a coupling for 
only a few locations around the simulated galaxy rather that at all locations.

A 3D phase-mixing simulation with an initial dispersion of $10$ \kms in $V_z$ was unable to reproduce the ridges in the $\langle V_z \rangle$\vphiR{} maps. This suggests that 
the origin of features in $V_z$ is dynamical with the self gravity of the disc playing a role. The simulations of both the unperturbed disc and the 
disc perturbed by a massive satellite show ridges in the $\langle V_z \rangle(R,V_{\phi})$ maps (\autoref{fig:multi_nb_ridge_compare}).  For the two cases of the perturbed disc, the profile of $z$ as a function of orbital energy is also found to show undulations (\autoref{fig:dens_correl}). For the case of the high-mass perturber, the most prominent peak in $\langle V_z\rangle$ shows a clear match with the most prominent peak in $\langle z\rangle$. While an interaction with 
an orbiting satellite can induce coupling between planar and vertical motions, e.g. they are known to generate warps, the case of an unperturbed disc generating such a coupling is intriguing. However, \citet{1997A&A...318..747M} have shown that {\it non-linear coupling between the Galactic spiral waves and the warp waves can lead to outer warps in isolated disc galaxies co-existent with corrugations (undulations) over the inner disc.} We propose to investigate this important insight further in the next paper.

Another example of coupling between 
the vertical and planar motion is the existence of 
the phase-spiral in the $(z,V_z)$ and $(z,V_R)$ planes \citep{2018Natur.561..360A,2019MNRAS.tmp..222B}. This phase-spiral is seen in density maps, $\langle V_R \rangle$ maps and $\langle V_{\phi} \rangle $ maps. Such a coupling can be generated by the impact of a satellite passing through the disc \citep[e.g.][]{bin18a}, or due to the buckling of the bar \citep[e.g.][]{2018arXiv181109205K}. {\it So far there has been no observation or simulation that suggests any link between the arches and ridges, and the $(z,V_z)$ phase-spiral.}

\begin{figure}  
\includegraphics[width=1.0\columnwidth]{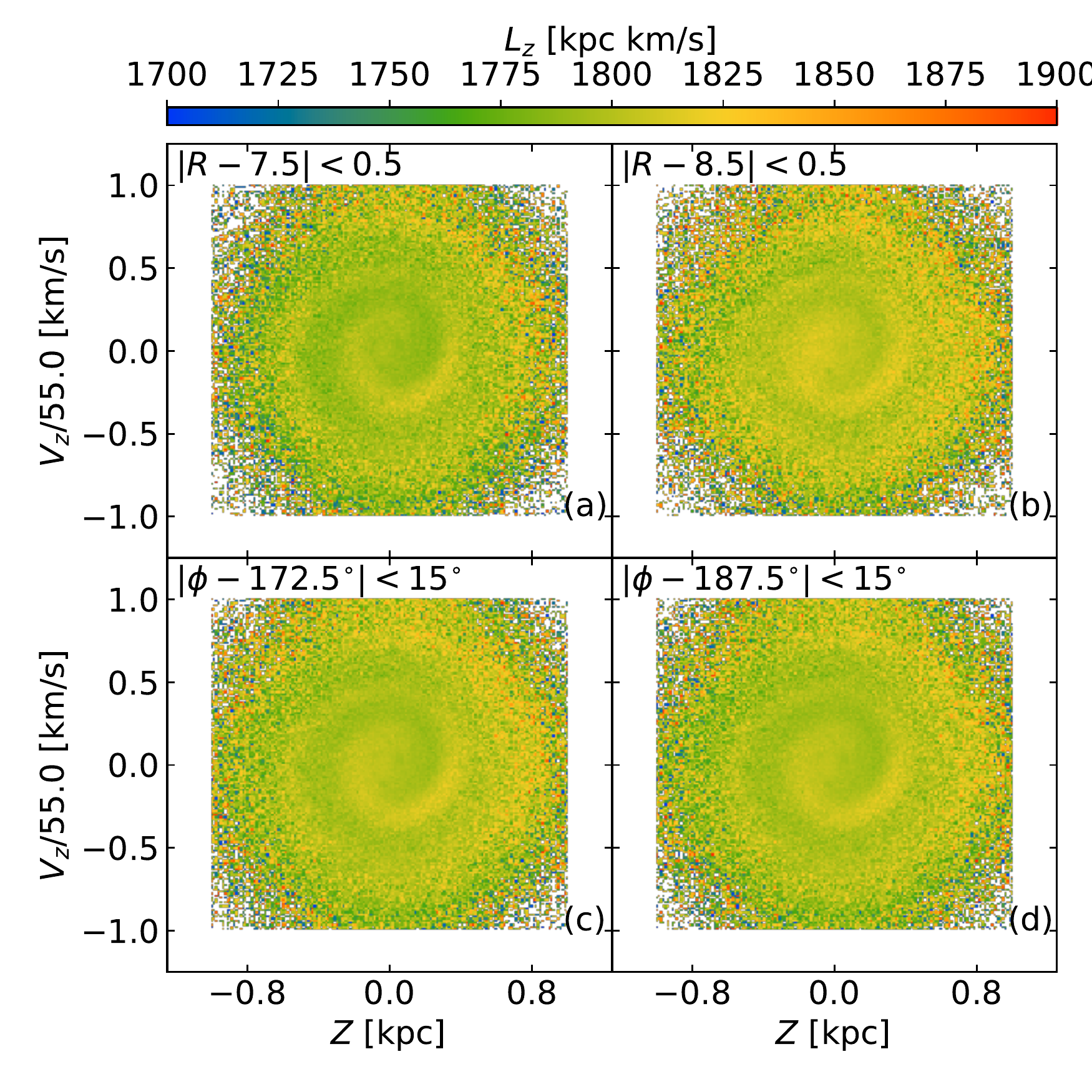}
\caption{Map of median $\langle L_z \rangle$ in the $(z,V_z)$ plane for various different Galactic positions. The orientation of the spiral remains unchanged with change in location.}
\label{fig:gaia_spiral1}
\end{figure}

\begin{figure*}  
\includegraphics[width=2.0\columnwidth]{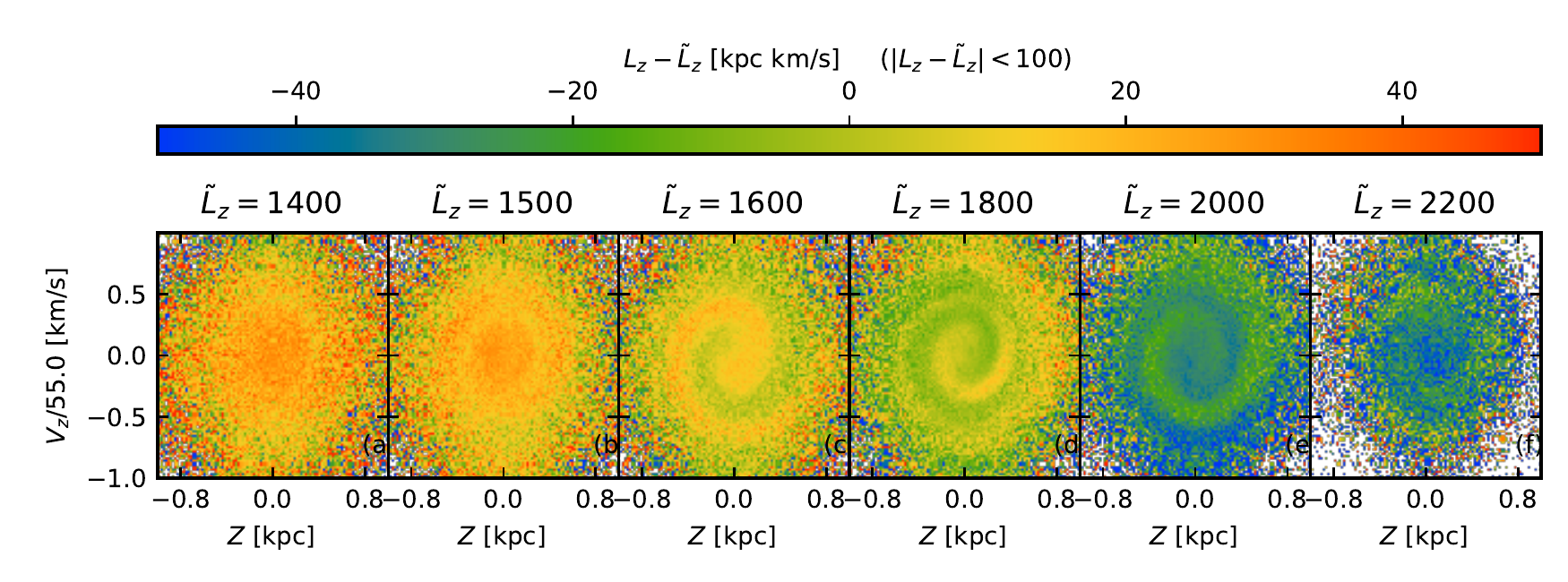}
\caption{The distribution of stars in the $(z,V_z)$ plane for 200 kpc \kms{} wide slices in angular momentum $L_z$. The overall median $L_z$ for each slice is labelled on the top. The image shows a map of median residual angular momentum in each pixel. The spiral pattern is visible for a wide range of angular momentum and 
the orientation of the spiral is found to change with $L_z$.}
\label{fig:gaia_spiral2}
\end{figure*}

We note that the average $z$ or $V_z$ integrated over a single ($z,V_z$) phase-space spiral is non-zero and depends upon the orientation of the spiral. So if the orientation 
of the spiral changes with $L_z$, we can expect a change of $\langle V_z\rangle$ with $L_z$.  
We find that $L_z$ is a more robust quantity to characterize the phase-space spiral compared to $V_{\phi}$. This is because the spiral pattern for a given $L_z$ (or orbital energy) is almost invariant with the 
Galactocentric radius $R$ (also with azimuth $\phi$; \autoref{fig:gaia_spiral1}), within a distance of around 1 kpc around the Sun, but $V_{\phi}$ is not. When the $(z,V_z)$ plane is studied for different 
values of $L_z$, we find that the spiral pattern is present for a wide range of $L_z$ and the orientation of the spiral changes with $L_z$ (\autoref{fig:gaia_spiral2}). However, the density distribution along the spiral is not constant and this can override any signatures in $\langle V_z\rangle$ generated by the spiral. Therefore, at this stage it is difficult to establish any link between the phase-space spiral and the ridges or the warp.

To conclude, there are many competing and interlocking dynamical processes occurring in the Galaxy. We have a bar, which leaves its imprint on the kinematics through resonances. We have multiple spiral arms, which are thought to be transient and can generate multiple features in kinematics. We have a warp which can couple planar and vertical motions. Other than the above 
mentioned internal mechanisms to excite the disc, there are also external mechanisms like interaction with orbiting satellites, e.g., Gaia--Enceladus, Sgr, LMC, SMC, and so on \citep{2018Natur.563...85H}. These can also couple vertical and 
planar motions. Clearly, it is important 
to understand and study the effect of these mechanisms individually. However, in future, we need to devise 
ways to study the different mechanisms together as 
the combined effect of mechanisms could be very different.

\graphicspath{{figures/}} 

\graphicspath{{figures/}} 

\section*{Acknowledgments}
We would like to thank the anonymous referee for their insightful comments.

The GALAH survey is based on observations made at the Australian Astronomical Observatory, under programmes A/2013B/13, A/2014A/25, A/2015A/19, A/2017A/18. We acknowledge the traditional owners of the land on which the AAT stands, the Gamilaraay people, and pay our respects to elders past and present. Parts of this research were conducted by the Australian Research Council Centre of Excellence for All Sky Astrophysics in 3 Dimensions (ASTRO 3D), through project number CE170100013.

This work has made use of data from the European Space Agency (ESA) mission
{\it Gaia} (\url{https://www.cosmos.esa.int/gaia}), processed by the {\it Gaia}
Data Processing and Analysis Consortium (DPAC,
\url{https://www.cosmos.esa.int/web/gaia/dpac/consortium}). Funding for the DPAC
has been provided by national institutions, in particular the institutions
participating in the {\it Gaia} Multilateral Agreement.

SK is supported by a Faculty of Science Postgraduate Scholarship at the University of Sydney.

JBH is supported by an ARC Australian Laureate Fellowship (FL140100278) and the ARC Centre of Excellence for All Sky Astrophysics in 3 Dimensions (ASTRO-3D) through project number CE170100013. SS is funded by a Dean's University Fellowship and through JBH's Laureate Fellowship, which also supports TTG and GDS. MJH is supported by an ASTRO-3D Fellowship. JK is supported by a Discovery Project grant from the Australian Research Council (DP150104667) awarded to JBH.

SB acknowledges funds from the Alexander von Humboldt Foundation in the framework of the Sofja Kovalevskaja Award endowed by the Federal Ministry of Education and Research.

HST-HF2-51425.001 awarded by the Space Telescope Science Institute.

JBH \& TTG acknowledge the Sydney Informatics Hub and the University of Sydney's high performance computing (HPC) cluster Artemis for providing the HPC resources that have contributed to the some of the research results reported within this paper. Parts of this project were undertaken with the assistance of resources and services from the National Computational Infrastructure (NCI), which is supported by the Australian Government.

This research has made use of Astropy, a community-developed core Python package for Astronomy \citep{2018AJ....156..123A}. This research has made use of NumPy (Walt et al., 2011), SciPy, and MatPlotLib (Hunter, 2007).



\bibliographystyle{mnras}
\interlinepenalty=10000
\bibliography{biblio.bib} 

\begin{thebibliography}{}
\makeatletter
\relax
\def\mn@urlcharsother{\let\do\@makeother \do\$\do\&\do\#\do\^\do\_\do\%\do\~}
\def\mn@doi{\begingroup\mn@urlcharsother \@ifnextchar [ {\mn@doi@}
  {\mn@doi@[]}}
\def\mn@doi@[#1]#2{\def\@tempa{#1}\ifx\@tempa\@empty \href
  {http://dx.doi.org/#2} {doi:#2}\else \href {http://dx.doi.org/#2} {#1}\fi
  \endgroup}
\def\mn@eprint#1#2{\mn@eprint@#1:#2::\@nil}
\def\mn@eprint@arXiv#1{\href {http://arxiv.org/abs/#1} {{\tt arXiv:#1}}}
\def\mn@eprint@dblp#1{\href {http://dblp.uni-trier.de/rec/bibtex/#1.xml}
  {dblp:#1}}
\def\mn@eprint@#1:#2:#3:#4\@nil{\def\@tempa {#1}\def\@tempb {#2}\def\@tempc
  {#3}\ifx \@tempc \@empty \let \@tempc \@tempb \let \@tempb \@tempa \fi \ifx
  \@tempb \@empty \def\@tempb {arXiv}\fi \@ifundefined
  {mn@eprint@\@tempb}{\@tempb:\@tempc}{\expandafter \expandafter \csname
  mn@eprint@\@tempb\endcsname \expandafter{\@tempc}}}

\bibitem[\protect\citeauthoryear{{Antoja} et~al.,}{{Antoja}
  et~al.}{2014}]{2014A&A...563A..60A}
{Antoja} T.,  et~al., 2014, \mn@doi [\aap] {10.1051/0004-6361/201322623}, \href
  {https://ui.adsabs.harvard.edu/\#abs/2014A&A...563A..60A} {563, A60}

\bibitem[\protect\citeauthoryear{{Antoja} et~al.,}{{Antoja}
  et~al.}{2018}]{2018Natur.561..360A}
{Antoja} T.,  et~al., 2018, \mn@doi [\nat] {10.1038/s41586-018-0510-7}, \href
  {http://adsabs.harvard.edu/abs/2018Natur.561..360A} {561, 360}

\bibitem[\protect\citeauthoryear{{Astropy Collaboration} et~al.,}{{Astropy
  Collaboration} et~al.}{2018}]{2018AJ....156..123A}
{Astropy Collaboration} et~al., 2018, \mn@doi [\aj] {10.3847/1538-3881/aabc4f},
  \href {https://ui.adsabs.harvard.edu/abs/2018AJ....156..123A} {156, 123}

\bibitem[\protect\citeauthoryear{{Bennett} \& {Bovy}}{{Bennett} \&
  {Bovy}}{2019}]{2019MNRAS.482.1417B}
{Bennett} M.,  {Bovy} J.,  2019, \mn@doi [\mnras] {10.1093/mnras/sty2813},
  \href {https://ui.adsabs.harvard.edu/\#abs/2019MNRAS.482.1417B} {482, 1417}

\bibitem[\protect\citeauthoryear{{Bensby}, {Feltzing}  \& {Oey}}{{Bensby}
  et~al.}{2014}]{2014A&A...562A..71B}
{Bensby} T.,  {Feltzing} S.,   {Oey} M.~S.,  2014, \mn@doi [\aap]
  {10.1051/0004-6361/201322631}, \href
  {https://ui.adsabs.harvard.edu/\#abs/2014A&A...562A..71B} {562, A71}

\bibitem[\protect\citeauthoryear{{Binney} \& {Sch\"onrich}}{{Binney} \&
  {Sch\"onrich}}{2018}]{bin18a}
{Binney} J.,  {Sch\"onrich} R.,  2018, \mnras, 481, 1501

\bibitem[\protect\citeauthoryear{{Bland-Hawthorn} \&
  {Gerhard}}{{Bland-Hawthorn} \& {Gerhard}}{2016}]{2016ARA&A..54..529B}
{Bland-Hawthorn} J.,  {Gerhard} O.,  2016, \mn@doi [\araa]
  {10.1146/annurev-astro-081915-023441}, \href
  {http://adsabs.harvard.edu/abs/2016ARA%26A..54..529B} {54, 529}

\bibitem[\protect\citeauthoryear{{Bland-Hawthorn} et~al.,}{{Bland-Hawthorn}
  et~al.}{2019}]{2019MNRAS.tmp..222B}
{Bland-Hawthorn} J.,  et~al., 2019, \mn@doi [\mnras] {10.1093/mnras/stz217}

\bibitem[\protect\citeauthoryear{{Bovy}}{{Bovy}}{2015}]{2015ApJS..216...29B}
{Bovy} J.,  2015, \mn@doi [\apjs] {10.1088/0067-0049/216/2/29}, \href
  {http://adsabs.harvard.edu/abs/2015ApJS..216...29B} {216, 29}

\bibitem[\protect\citeauthoryear{{Bovy} \& {Rix}}{{Bovy} \&
  {Rix}}{2013}]{2013ApJ...779..115B}
{Bovy} J.,  {Rix} H.-W.,  2013, \mn@doi [\apj] {10.1088/0004-637X/779/2/115},
  \href {https://ui.adsabs.harvard.edu/abs/2013ApJ...779..115B} {779, 115}

\bibitem[\protect\citeauthoryear{{Buder} et~al.,}{{Buder}
  et~al.}{2018}]{2018MNRAS.478.4513B}
{Buder} S.,  et~al., 2018, \mn@doi [\mnras] {10.1093/mnras/sty1281}, \href
  {https://ui.adsabs.harvard.edu/\#abs/2018MNRAS.478.4513B} {478, 4513}

\bibitem[\protect\citeauthoryear{{Chequers}, {Widrow}  \& {Darling}}{{Chequers}
  et~al.}{2018}]{2018MNRAS.480.4244C}
{Chequers} M.~H.,  {Widrow} L.~M.,   {Darling} K.,  2018, \mn@doi [\mnras]
  {10.1093/mnras/sty2114}, \href
  {https://ui.adsabs.harvard.edu/abs/2018MNRAS.480.4244C} {480, 4244}

\bibitem[\protect\citeauthoryear{{D'Onghia}, {Madau}, {Vera-Ciro}, {Quillen}
  \& {Hernquist}}{{D'Onghia} et~al.}{2016}]{2016ApJ...823....4D}
{D'Onghia} E.,  {Madau} P.,  {Vera-Ciro} C.,  {Quillen} A.,   {Hernquist} L.,
  2016, \mn@doi [\apj] {10.3847/0004-637X/823/1/4}, \href
  {http://adsabs.harvard.edu/abs/2016ApJ...823....4D} {823, 4}

\bibitem[\protect\citeauthoryear{{Dehnen}}{{Dehnen}}{2000}]{2000AJ....119..800D}
{Dehnen} W.,  2000, \mn@doi [\aj] {10.1086/301226}, \href
  {http://adsabs.harvard.edu/abs/2000AJ....119..800D} {119, 800}

\bibitem[\protect\citeauthoryear{{Duong} et~al.,}{{Duong}
  et~al.}{2018}]{2018MNRAS.476.5216D}
{Duong} L.,  et~al., 2018, \mn@doi [\mnras] {10.1093/mnras/sty525}, \href
  {https://ui.adsabs.harvard.edu/abs/2018MNRAS.476.5216D} {476, 5216}

\bibitem[\protect\citeauthoryear{{Flynn}, {Sommer-Larsen}  \&
  {Christensen}}{{Flynn} et~al.}{1996}]{1996MNRAS.281.1027F}
{Flynn} C.,  {Sommer-Larsen} J.,   {Christensen} P.~R.,  1996, \mn@doi [\mnras]
  {10.1093/mnras/281.3.1027}, \href
  {https://ui.adsabs.harvard.edu/abs/1996MNRAS.281.1027F} {281, 1027}

\bibitem[\protect\citeauthoryear{{Fragkoudi} et~al.,}{{Fragkoudi}
  et~al.}{2019}]{2019arXiv190107568F}
{Fragkoudi} F.,  et~al., 2019, arXiv e-prints, \href
  {https://ui.adsabs.harvard.edu/\#abs/2019arXiv190107568F} {p.
  arXiv:1901.07568}

\bibitem[\protect\citeauthoryear{{Gaia Collaboration} et~al.,}{{Gaia
  Collaboration} et~al.}{2018a}]{2018A&A...616A...1G}
{Gaia Collaboration} et~al., 2018a, \mn@doi [\aap]
  {10.1051/0004-6361/201833051}, \href
  {http://adsabs.harvard.edu/abs/2018A%26A...616A...1G} {616, A1}

\bibitem[\protect\citeauthoryear{{Gaia Collaboration} et~al.,}{{Gaia
  Collaboration} et~al.}{2018b}]{2018A&A...616A..11G}
{Gaia Collaboration} et~al., 2018b, \mn@doi [\aap]
  {10.1051/0004-6361/201832865}, \href
  {https://ui.adsabs.harvard.edu/\#abs/2018A&A...616A..11G} {616, A11}

\bibitem[\protect\citeauthoryear{{G{\'o}mez} et~al.,}{{G{\'o}mez}
  et~al.}{2012}]{2012MNRAS.423.3727G}
{G{\'o}mez} F.~A.,  et~al., 2012, \mn@doi [\mnras]
  {10.1111/j.1365-2966.2012.21176.x}, \href
  {https://ui.adsabs.harvard.edu/\#abs/2012MNRAS.423.3727G} {423, 3727}

\bibitem[\protect\citeauthoryear{{G{\'o}mez}, {Minchev}, {O'Shea}, {Beers},
  {Bullock}  \& {Purcell}}{{G{\'o}mez} et~al.}{2013}]{2013MNRAS.429..159G}
{G{\'o}mez} F.~A.,  {Minchev} I.,  {O'Shea} B.~W.,  {Beers} T.~C.,  {Bullock}
  J.~S.,   {Purcell} C.~W.,  2013, \mn@doi [\mnras] {10.1093/mnras/sts327},
  \href {https://ui.adsabs.harvard.edu/\#abs/2013MNRAS.429..159G} {429, 159}

\bibitem[\protect\citeauthoryear{{Gravity Collaboration} et~al.,}{{Gravity
  Collaboration} et~al.}{2018}]{2018A&A...615L..15G}
{Gravity Collaboration} et~al., 2018, \mn@doi [\aap]
  {10.1051/0004-6361/201833718}, \href
  {https://ui.adsabs.harvard.edu/\#abs/2018A&A...615L..15G} {615, L15}

\bibitem[\protect\citeauthoryear{{Hattori}, {Gouda}, {Tagawa}, {Sakai}, {Yano},
  {Baba}  \& {Kumamoto}}{{Hattori} et~al.}{2019}]{2019MNRAS.484.4540H}
{Hattori} K.,  {Gouda} N.,  {Tagawa} H.,  {Sakai} N.,  {Yano} T.,  {Baba} J.,
  {Kumamoto} J.,  2019, \mn@doi [\mnras] {10.1093/mnras/stz266}, \href
  {https://ui.adsabs.harvard.edu/\#abs/2019MNRAS.484.4540H} {484, 4540}

\bibitem[\protect\citeauthoryear{{Hayden} et~al.,}{{Hayden}
  et~al.}{2014}]{2014AJ....147..116H}
{Hayden} M.~R.,  et~al., 2014, \mn@doi [\aj] {10.1088/0004-6256/147/5/116},
  \href {http://adsabs.harvard.edu/abs/2014AJ....147..116H} {147, 116}

\bibitem[\protect\citeauthoryear{{Helmi}, {Babusiaux}, {Koppelman}, {Massari},
  {Veljanoski}  \& {Brown}}{{Helmi} et~al.}{2018}]{2018Natur.563...85H}
{Helmi} A.,  {Babusiaux} C.,  {Koppelman} H.~H.,  {Massari} D.,  {Veljanoski}
  J.,   {Brown} A. G.~A.,  2018, \mn@doi [\nat] {10.1038/s41586-018-0625-x},
  \href {https://ui.adsabs.harvard.edu/\#abs/2018Natur.563...85H} {563, 85}

\bibitem[\protect\citeauthoryear{{Hernquist}}{{Hernquist}}{1990}]{her90a}
{Hernquist} L.,  1990, \mn@doi [\apj] {10.1086/168845}, \href
  {http://adsabs.harvard.edu/abs/1990ApJ...356..359H} {356, 359}

\bibitem[\protect\citeauthoryear{{Huang} et~al.,}{{Huang}
  et~al.}{2018}]{2018ApJ...864..129H}
{Huang} Y.,  et~al., 2018, \mn@doi [\apj] {10.3847/1538-4357/aad285}, \href
  {https://ui.adsabs.harvard.edu/\#abs/2018ApJ...864..129H} {864, 129}

\bibitem[\protect\citeauthoryear{{Hunt} \& {Bovy}}{{Hunt} \&
  {Bovy}}{2018}]{2018MNRAS.477.3945H}
{Hunt} J. A.~S.,  {Bovy} J.,  2018, \mn@doi [\mnras] {10.1093/mnras/sty921},
  \href {https://ui.adsabs.harvard.edu/\#abs/2018MNRAS.477.3945H} {477, 3945}

\bibitem[\protect\citeauthoryear{{Hunt}, {Hong}, {Bovy}, {Kawata}  \&
  {Grand}}{{Hunt} et~al.}{2018}]{2018MNRAS.481.3794H}
{Hunt} J.~A.~S.,  {Hong} J.,  {Bovy} J.,  {Kawata} D.,   {Grand} R.~J.~J.,
  2018, \mn@doi [\mnras] {10.1093/mnras/sty2532}, \href
  {http://adsabs.harvard.edu/abs/2018MNRAS.481.3794H} {481, 3794}

\bibitem[\protect\citeauthoryear{{Kafle}, {Sharma}, {Lewis}  \&
  {Bland-Hawthorn}}{{Kafle} et~al.}{2014}]{2014ApJ...794...59K}
{Kafle} P.~R.,  {Sharma} S.,  {Lewis} G.~F.,   {Bland-Hawthorn} J.,  2014,
  \mn@doi [\apj] {10.1088/0004-637X/794/1/59}, \href
  {http://adsabs.harvard.edu/abs/2014ApJ...794...59K} {794, 59}

\bibitem[\protect\citeauthoryear{{Kawata}, {Baba}, {Ciuc{\v{a}}}, {Cropper},
  {Grand}, {Hunt}  \& {Seabroke}}{{Kawata} et~al.}{2018}]{2018MNRAS.479L.108K}
{Kawata} D.,  {Baba} J.,  {Ciuc{\v{a}}} I.,  {Cropper} M.,  {Grand} R. J.~J.,
  {Hunt} J. A.~S.,   {Seabroke} G.,  2018, \mn@doi [\mnras]
  {10.1093/mnrasl/sly107}, \href
  {https://ui.adsabs.harvard.edu/\#abs/2018MNRAS.479L.108K} {479, L108}

\bibitem[\protect\citeauthoryear{{Kazantzidis}, {Magorrian}  \&
  {Moore}}{{Kazantzidis} et~al.}{2004}]{kaz04a}
{Kazantzidis} S.,  {Magorrian} J.,   {Moore} B.,  2004, \mn@doi [\apj]
  {10.1086/380192}, \href {http://adsabs.harvard.edu/abs/2004ApJ...601...37K}
  {601, 37}

\bibitem[\protect\citeauthoryear{{Khoperskov}, {Di Matteo}, {Gerhard}, {Katz},
  {Haywood}, {Combes}, {Berczik}  \& {Gomez}}{{Khoperskov}
  et~al.}{2018}]{2018arXiv181109205K}
{Khoperskov} S.,  {Di Matteo} P.,  {Gerhard} O.,  {Katz} D.,  {Haywood} M.,
  {Combes} F.,  {Berczik} P.,   {Gomez} A.,  2018, arXiv e-prints, \href
  {https://ui.adsabs.harvard.edu/\#abs/2018arXiv181109205K} {p.
  arXiv:1811.09205}

\bibitem[\protect\citeauthoryear{{Laporte}, {Minchev}, {Johnston}  \&
  {G{\'o}mez}}{{Laporte} et~al.}{2019}]{2019MNRAS.485.3134L}
{Laporte} C. F.~P.,  {Minchev} I.,  {Johnston} K.~V.,   {G{\'o}mez} F.~A.,
  2019, \mn@doi [\mnras] {10.1093/mnras/stz583}, \href
  {https://ui.adsabs.harvard.edu/abs/2019MNRAS.485.3134L} {485, 3134}

\bibitem[\protect\citeauthoryear{{Luri} et~al.,}{{Luri}
  et~al.}{2018}]{2018A&A...616A...9L}
{Luri} X.,  et~al., 2018, \mn@doi [\aap] {10.1051/0004-6361/201832964}, \href
  {http://adsabs.harvard.edu/abs/2018A%26A...616A...9L} {616, A9}

\bibitem[\protect\citeauthoryear{{Mackereth} et~al.,}{{Mackereth}
  et~al.}{2017}]{2017MNRAS.471.3057M}
{Mackereth} J.~T.,  et~al., 2017, \mn@doi [\mnras] {10.1093/mnras/stx1774},
  \href {https://ui.adsabs.harvard.edu/\#abs/2017MNRAS.471.3057M} {471, 3057}

\bibitem[\protect\citeauthoryear{{Malhan}, {Ibata}  \& {Martin}}{{Malhan}
  et~al.}{2018}]{2018MNRAS.481.3442M}
{Malhan} K.,  {Ibata} R.~A.,   {Martin} N.~F.,  2018, \mn@doi [\mnras]
  {10.1093/mnras/sty2474}, \href
  {https://ui.adsabs.harvard.edu/\#abs/2018MNRAS.481.3442M} {481, 3442}

\bibitem[\protect\citeauthoryear{{Masset} \& {Tagger}}{{Masset} \&
  {Tagger}}{1997}]{1997A&A...318..747M}
{Masset} F.,  {Tagger} M.,  1997, \aap, \href
  {http://adsabs.harvard.edu/abs/1997A%26A...318..747M} {318, 747}

\bibitem[\protect\citeauthoryear{{Minchev}, {Quillen}, {Williams}, {Freeman},
  {Nordhaus}, {Siebert}  \& {Bienaym{\'e}}}{{Minchev}
  et~al.}{2009}]{2009MNRAS.396L..56M}
{Minchev} I.,  {Quillen} A.~C.,  {Williams} M.,  {Freeman} K.~C.,  {Nordhaus}
  J.,  {Siebert} A.,   {Bienaym{\'e}} O.,  2009, \mn@doi [\mnras]
  {10.1111/j.1745-3933.2009.00661.x}, \href
  {http://adsabs.harvard.edu/abs/2009MNRAS.396L..56M} {396, L56}

\bibitem[\protect\citeauthoryear{{Miyamoto} \& {Nagai}}{{Miyamoto} \&
  {Nagai}}{1975a}]{miy75a}
{Miyamoto} M.,  {Nagai} R.,  1975a, \pasj, \href
  {http://adsabs.harvard.edu/abs/1975PASJ...27..533M} {27, 533}

\bibitem[\protect\citeauthoryear{{Miyamoto} \& {Nagai}}{{Miyamoto} \&
  {Nagai}}{1975b}]{1975PASJ...27..533M}
{Miyamoto} M.,  {Nagai} R.,  1975b, \pasj, \href
  {https://ui.adsabs.harvard.edu/abs/1975PASJ...27..533M} {27, 533}

\bibitem[\protect\citeauthoryear{{Monari}, {Famaey}, {Fouvry}  \&
  {Binney}}{{Monari} et~al.}{2017}]{2017MNRAS.471.4314M}
{Monari} G.,  {Famaey} B.,  {Fouvry} J.-B.,   {Binney} J.,  2017, \mn@doi
  [\mnras] {10.1093/mnras/stx1825}, \href
  {https://ui.adsabs.harvard.edu/\#abs/2017MNRAS.471.4314M} {471, 4314}

\bibitem[\protect\citeauthoryear{{Niederste-Ostholt}, {Belokurov}, {Evans}  \&
  {Peñarrubia}}{{Niederste-Ostholt} et~al.}{2010}]{nie10a}
{Niederste-Ostholt} M.,  {Belokurov} V.,  {Evans} N.~W.,   {Peñarrubia} J.,
  2010, \apj, 712, 516

\bibitem[\protect\citeauthoryear{{Nieva} \& {Przybilla}}{{Nieva} \&
  {Przybilla}}{2012}]{2012A&A...539A.143N}
{Nieva} M.-F.,  {Przybilla} N.,  2012, \mn@doi [\aap]
  {10.1051/0004-6361/201118158}, \href
  {http://adsabs.harvard.edu/abs/2012A%26A...539A.143N} {539, A143}

\bibitem[\protect\citeauthoryear{{P{\'e}rez-Villegas}, {Portail}, {Wegg}  \&
  {Gerhard}}{{P{\'e}rez-Villegas} et~al.}{2017}]{2017ApJ...840L...2P}
{P{\'e}rez-Villegas} A.,  {Portail} M.,  {Wegg} C.,   {Gerhard} O.,  2017,
  \mn@doi [\apj] {10.3847/2041-8213/aa6c26}, \href
  {https://ui.adsabs.harvard.edu/\#abs/2017ApJ...840L...2P} {840, L2}

\bibitem[\protect\citeauthoryear{{Perret}, {Renaud}, {Epinat}, {Amram},
  {Bournaud}, {Contini}, {Teyssier}  \& {Lambert}}{{Perret}
  et~al.}{2014}]{per14c}
{Perret} V.,  {Renaud} F.,  {Epinat} B.,  {Amram} P.,  {Bournaud} F.,
  {Contini} T.,  {Teyssier} R.,   {Lambert} J.-C.,  2014, \mn@doi [\aap]
  {10.1051/0004-6361/201322395}, \href
  {http://adsabs.harvard.edu/abs/2014A%26A...562A...1P} {562, A1}

\bibitem[\protect\citeauthoryear{{Perryman} et~al.,}{{Perryman}
  et~al.}{1997}]{1997A&A...323L..49P}
{Perryman} M.~A.~C.,  et~al., 1997, \aap, \href
  {http://adsabs.harvard.edu/abs/1997A%26A...323L..49P} {323, L49}

\bibitem[\protect\citeauthoryear{{Poggio}, {Drimmel}, {Smart}, {Spagna}  \&
  {Lattanzi}}{{Poggio} et~al.}{2017}]{2017A&A...601A.115P}
{Poggio} E.,  {Drimmel} R.,  {Smart} R.~L.,  {Spagna} A.,   {Lattanzi} M.~G.,
  2017, \mn@doi [\aap] {10.1051/0004-6361/201629916}, \href
  {http://adsabs.harvard.edu/abs/2017A%26A...601A.115P} {601, A115}

\bibitem[\protect\citeauthoryear{{Poggio} et~al.,}{{Poggio}
  et~al.}{2018}]{2018MNRAS.481L..21P}
{Poggio} E.,  et~al., 2018, \mn@doi [\mnras] {10.1093/mnrasl/sly148}, \href
  {http://adsabs.harvard.edu/abs/2018MNRAS.481L..21P} {481, L21}

\bibitem[\protect\citeauthoryear{{Price-Whelan} \& {Bonaca}}{{Price-Whelan} \&
  {Bonaca}}{2018}]{2018ApJ...863L..20P}
{Price-Whelan} A.~M.,  {Bonaca} A.,  2018, \mn@doi [\apj]
  {10.3847/2041-8213/aad7b5}, \href
  {https://ui.adsabs.harvard.edu/\#abs/2018ApJ...863L..20P} {863, L20}

\bibitem[\protect\citeauthoryear{{Quillen}, {Dougherty}, {Bagley}, {Minchev}
  \& {Comparetta}}{{Quillen} et~al.}{2011}]{2011MNRAS.417..762Q}
{Quillen} A.~C.,  {Dougherty} J.,  {Bagley} M.~B.,  {Minchev} I.,
  {Comparetta} J.,  2011, \mn@doi [\mnras] {10.1111/j.1365-2966.2011.19349.x},
  \href {http://adsabs.harvard.edu/abs/2011MNRAS.417..762Q} {417, 762}

\bibitem[\protect\citeauthoryear{{Quillen} et~al.,}{{Quillen}
  et~al.}{2018}]{2018MNRAS.480.3132Q}
{Quillen} A.~C.,  et~al., 2018, \mn@doi [\mnras] {10.1093/mnras/sty2077}, \href
  {http://adsabs.harvard.edu/abs/2018MNRAS.480.3132Q} {480, 3132}

\bibitem[\protect\citeauthoryear{{Ramos}, {Antoja}  \& {Figueras}}{{Ramos}
  et~al.}{2018}]{2018A&A...619A..72R}
{Ramos} P.,  {Antoja} T.,   {Figueras} F.,  2018, \mn@doi [\aap]
  {10.1051/0004-6361/201833494}, \href
  {https://ui.adsabs.harvard.edu/\#abs/2018A&A...619A..72R} {619, A72}

\bibitem[\protect\citeauthoryear{{Reid} \& {Brunthaler}}{{Reid} \&
  {Brunthaler}}{2004}]{2004ApJ...616..872R}
{Reid} M.~J.,  {Brunthaler} A.,  2004, \mn@doi [\apj] {10.1086/424960}, \href
  {http://adsabs.harvard.edu/abs/2004ApJ...616..872R} {616, 872}

\bibitem[\protect\citeauthoryear{{Sch{\"o}nrich} \& {Dehnen}}{{Sch{\"o}nrich}
  \& {Dehnen}}{2018}]{2018MNRAS.478.3809S}
{Sch{\"o}nrich} R.,  {Dehnen} W.,  2018, \mn@doi [\mnras]
  {10.1093/mnras/sty1256}, \href
  {http://adsabs.harvard.edu/abs/2018MNRAS.478.3809S} {478, 3809}

\bibitem[\protect\citeauthoryear{{Sch{\"o}nrich}, {Binney}  \&
  {Dehnen}}{{Sch{\"o}nrich} et~al.}{2010}]{2010MNRAS.403.1829S}
{Sch{\"o}nrich} R.,  {Binney} J.,   {Dehnen} W.,  2010, \mn@doi [\mnras]
  {10.1111/j.1365-2966.2010.16253.x}, \href
  {http://adsabs.harvard.edu/abs/2010MNRAS.403.1829S} {403, 1829}

\bibitem[\protect\citeauthoryear{{Sellwood}}{{Sellwood}}{2011}]{2011MNRAS.410.1637S}
{Sellwood} J.~A.,  2011, \mn@doi [\mnras] {10.1111/j.1365-2966.2010.17545.x},
  \href {http://adsabs.harvard.edu/abs/2011MNRAS.410.1637S} {410, 1637}

\bibitem[\protect\citeauthoryear{{Sellwood}}{{Sellwood}}{2012}]{2012ApJ...751...44S}
{Sellwood} J.~A.,  2012, \mn@doi [\apj] {10.1088/0004-637X/751/1/44}, \href
  {https://ui.adsabs.harvard.edu/\#abs/2012ApJ...751...44S} {751, 44}

\bibitem[\protect\citeauthoryear{{Sharma} et~al.,}{{Sharma}
  et~al.}{2018}]{2018MNRAS.473.2004S}
{Sharma} S.,  et~al., 2018, \mn@doi [\mnras] {10.1093/mnras/stx2582}, \href
  {http://adsabs.harvard.edu/abs/2018MNRAS.473.2004S} {473, 2004}

\bibitem[\protect\citeauthoryear{{Smith}, {Flynn}, {Candlish}, {Fellhauer}  \&
  {Gibson}}{{Smith} et~al.}{2015}]{2015MNRAS.448.2934S}
{Smith} R.,  {Flynn} C.,  {Candlish} G.~N.,  {Fellhauer} M.,   {Gibson} B.~K.,
  2015, \mn@doi [\mnras] {10.1093/mnras/stv228}, \href
  {https://ui.adsabs.harvard.edu/abs/2015MNRAS.448.2934S} {448, 2934}

\bibitem[\protect\citeauthoryear{{Soubiran} et~al.,}{{Soubiran}
  et~al.}{2018}]{2018A&A...616A...7S}
{Soubiran} C.,  et~al., 2018, \mn@doi [\aap] {10.1051/0004-6361/201832795},
  \href {https://ui.adsabs.harvard.edu/\#abs/2018A&A...616A...7S} {616, A7}

\bibitem[\protect\citeauthoryear{{Springel}, {Di Matteo}  \&
  {Hernquist}}{{Springel} et~al.}{2005}]{spr05c}
{Springel} V.,  {Di Matteo} T.,   {Hernquist} L.,  2005, \mn@doi [\mnras]
  {10.1111/j.1365-2966.2005.09238.x}, \href
  {http://adsabs.harvard.edu/abs/2005MNRAS.361..776S} {361, 776}

\bibitem[\protect\citeauthoryear{{Tepper-Garc{\'\i}a} \&
  {Bland-Hawthorn}}{{Tepper-Garc{\'\i}a} \& {Bland-Hawthorn}}{2018}]{tep18b}
{Tepper-Garc{\'\i}a} T.,  {Bland-Hawthorn} J.,  2018, \mnras, 478, 5263

\bibitem[\protect\citeauthoryear{{Teyssier}}{{Teyssier}}{2002}]{tey02a}
{Teyssier} R.,  2002, \mn@doi [\aap] {10.1051/0004-6361:20011817}, \href
  {http://adsabs.harvard.edu/abs/2002A%26A...385..337T} {385, 337}

\bibitem[\protect\citeauthoryear{{Trick}, {Coronado}  \& {Rix}}{{Trick}
  et~al.}{2019}]{2019MNRAS.484.3291T}
{Trick} W.~H.,  {Coronado} J.,   {Rix} H.-W.,  2019, \mn@doi [\mnras]
  {10.1093/mnras/stz209}, \href
  {https://ui.adsabs.harvard.edu/abs/2019MNRAS.484.3291T} {484, 3291}

\bibitem[\protect\citeauthoryear{{Wittenmyer} et~al.,}{{Wittenmyer}
  et~al.}{2018}]{2018AJ....155...84W}
{Wittenmyer} R.~A.,  et~al., 2018, \mn@doi [\aj] {10.3847/1538-3881/aaa3e4},
  \href {http://adsabs.harvard.edu/abs/2018AJ....155...84W} {155, 84}

\bibitem[\protect\citeauthoryear{{Zwitter} et~al.,}{{Zwitter}
  et~al.}{2018}]{2018MNRAS.481..645Z}
{Zwitter} T.,  et~al., 2018, \mn@doi [\mnras] {10.1093/mnras/sty2293}, \href
  {https://ui.adsabs.harvard.edu/\#abs/2018MNRAS.481..645Z} {481, 645}

\makeatother
\end{thebibliography}


\appendix
\section{Gaia SQL query}
\label{app:gaia_sql}
\begin{verbatim}
Select * from gaiadr2.gaia_source G
where G.parallax IS NOT Null
AND G.parallax_error/G.parallax < 0.2
AND G.parallax > 0.\
where G.radial_velocity IS NOT Null
\end{verbatim}



\bsp	
\label{lastpage}
\end{document}